\documentclass[prd,aps,nofootinbib,10pt]{revtex4}
\usepackage{amsmath,graphicx,color,epsfig}

\begin{document}
\title{ Transition form factors of $B$ decays into p-wave axial-vector mesons in the perturbative QCD approach}
\author{ Run-Hui Li$^{a,b}$, Cai-Dian L\"u$^{a,c}$ and  Wei Wang$^{a}$ }

\affiliation{
 \it $^a$ Institute of High Energy Physics, P.O. Box 918(4) Beijing 100049, Peoples' Republic of  China\\
 \it $^b$  School of Physics, Shandong University, Jinan 250100, Peoples' Republic of China \\
 \it $^c$ Kavli Institute for Theoretical Physics, Peoples' Republic of  China}

\begin{abstract}
The $B_{u,d,s}\to V,A$ form factors are studied in perturbative QCD
approach ($V,A$ denote a vector meson and two kinds of p-wave
axial-vector mesons: $^3P_1$ and $^1P_1$ states, respectively.). The
form factors are directly studied in the large recoiling region and
extrapolated to the whole kinematic region within the dipole
parametrization. Adopting decay constants with different signs for
the two kinds of axial-vectors, we find that the two kinds of $B\to
A$ form factors have the same sign.  The two strange mesons $K_{1A}$
and $K_{1B}$ mix with each other via the SU(3) symmetry breaking
effect. In order to reduce the ambiguities in the mixing angle
between $K_{1A}$ and $K_{1B}$, we propose a model-independent way
that utilizes the B decay data.  Most of the branching fractions of
the semilteptonic $B\to Al\bar \nu_l$ decays are of the order
$10^{-4}$, which still need experimental tests in the on-going and
forthcoming experiments.
\end{abstract}
\maketitle
%%%%%%%%%%%%%%%%%%%%%%%%%%%%%%%%%%%%%%%%%%%%%%%%%%%%%%%%%%%%
\section{Introduction}\label{section:introduction}
%%%%%%%%%%%%%%%%%%%%%%%%%%%%%%%%%%%%%%%%%%%%%%%%%%%%%%%%%%%%

In rare charmless $B$ decays, the main experimental observables are
branching ratios and CP asymmetries. To predict these quantities,
one needs to compute the hadronic decay amplitudes. Since
hadronizations are involved in these decay channels,  predictions on
these observables are often polluted by our poor knowledge of the
non-perturbative QCD. But fortunately, it has been shown that in
$m_b\to \infty$ limit, the decay amplitudes are under control. For
example, if the recoiling meson in the final state moves very fast,
a hard gluon is required to kick the soft light quark in B meson
into a collinear one and then the process is calculable. Keeping
quarks' transverse momentum, the perturbative QCD (PQCD) approach
\cite{Keum:2000ph} is free of endpoint divergence and the Sudakov
formalism makes it more self-consistent. A bigger advantage is that
we can really do the form factor calculation and the quantitative
annihilation type diagram calculation in this approach. The
importance of annihilation diagrams is already tested in the
predictions of direct CP asymmetries of $B^0 \to \pi^+\pi^-$,
$K^+\pi^-$ decays \cite{Keum:2000ph,direct} and in the explanation
of $B\to \phi K^*$ polarization problem \cite{kphi,kphi-e}.

In the quark model, the possible quantum numbers $J^{PC}$ for the
orbitally excited axial-vector mesons are $1^{++}$ or $1^{+-}$,
depending on different spin couplings of the two quarks. In the
SU(3) limit, these mesons can not mix with each other; but since the
$s$ quark is heavier than $u,d$ quarks, $K_1(1270)$ and $K_1(1400)$
are not purely $1^3P_1$ or $1^1P_1$ states. These two mesons are
believed to be mixtures of $K_{1A}$ and $K_{1B}$, where $K_{1A}$ and
$K_{1B}$ are $^3P_1$ and $^1P_1$ states, respectively. Analogous to
$\eta$ and $\eta^\prime$, the flavor-singlet and flavor-octet
axial-vector meson can also mix with each other. In general, the
mixing angles can be determined by experimental data, but
unfortunately, there is not too much data on these mesons which
leaves the mixing angles much free. The $B$ meson decays offer a
promising opportunity to investigate these axial-vector mesons.
Since the observation of the $B\to J/\psi K_1$~\cite{Abe:2001wa} and
$D^* a_1(1260)$~\cite{Aubert:2002sp} decays, there are more and more
experimental studies on B meson decays involving a p-wave
axial-vector meson in the final state~\cite{Barberio:2006bi}. In the
present work, we use the PQCD approach to study the $B\to A$ form
factors and semileptonic $B\to Al\bar\nu$ decays. As a byproduct, we
also update the predictions on $B\to V$ form factors in the PQCD
approach. In the large recoiling region,  the $B\to A$ form factors
are directly predicted using the most recent inputs evaluated in the
QCD sum rules~\cite{Yang:2005gk,Yang:2007zt}. We also extrapolate
the form factors to  the whole kinematic region by adopting the
dipole parametrization to investigate the semileptonic $B\to
Al\bar\nu$ decays. Using the $\bar B^0\to D^+K_{1A}$ and $\bar
B^0\to D^+\pi^-$ decays, we also propose a model-independent method
to remove the ambiguity in the mixing between the two strange
axial-vector mesons.

This paper is organized as follows: In section II, we give the input
quantities, including wave function of the $B$-meson, light-cone
distribution amplitudes of the light vector mesons and light
axial-vector mesons and input values of the various mesonic decay
constants. In section~\ref{section:formfactor}, we give the
factorization formulae and the numerical results  for the $B\to V$
and $B\to A$ form factors, discuss the mixing between the strange
axial-vector mesons and make the predictions on the semileptonic
$B\to Al\bar\nu_l$ decays. Our summary is given in the last section.
Appendix A contains various functions that enter the factorization
formulae in the PQCD approach.

%%%%%%%%%%%%%%%%%%%%%%%%%%%%%%%%%%%%%%%%%%%%%%%%%%%%%%%%%%
\section{Formalism of the PQCD approach and Inputs}\label{section:pqcd}
%%%%%%%%%%%%%%%%%%%%%%%%%%%%%%%%%%%%%%%%%%%%%%%%%%%%%%%%%%
%
%
%%%%%%%%%%%%%%%%%%%%%%%%%%%%%%%%%%%%%%%%%%%%%%%%%%%%%%%%%%%
%\subsection{Notations and conventions}
%%%%%%%%%%%%%%%%%%%%%%%%%%%%%%%%%%%%%%%%%%%%%%%%%%%%%%%%%%

%
%
%%%%%%%%%%%%%%%%%%%%%%%%%%%%%%%%%%%%%%%%%%%%%%%%%%%%%%%%%%%
%\subsection{Wave functions of $B$ mesons}
%%%%%%%%%%%%%%%%%%%%%%%%%%%%%%%%%%%%%%%%%%%%%%%%%%%%%%%%%%%

We will work in the rest frame of the B meson and use light-cone
coordinates. In the heavy quark limit the mass difference between b
quark and B meson is negligible: $m_b\simeq m_B$. Masses of
axial-vector mesons are very small compared with the b quark mass,
we keep them up to the first order. Since the
light(vector/axial-vector) meson in the final state moves very fast
in the large-recoil region, we define its momentum mainly on the
plus direction in the light-cone coordinates. The momentum of B
meson and light mesons can be denoted  as
 \begin{eqnarray}
 P_{B_{(s)}}=\frac{m_{B_{(s)}}}{\sqrt{2}}(1,1,0_{\perp})\;,\;
 P_2=\frac{m_{B_{(s)}}}{\sqrt{2}}(\eta,\frac{r_2^2}{\eta},0_{\perp})\;,\label{eq:momentum}
 \end{eqnarray}
where $r_2\equiv \frac{m_{V/A}}{m_{B_{(s)}}}$, with $m_{V/A}$ as the
mass of the vector or axial-vector meson. For the momentum transfer
$q=P_{B_{(s)}}-P_2$, there exists $\eta\approx1-q^2/m_{B_{(s)}}^2$.
The momentum of the light antiquark in $B_{(s)}$ meson and the quark
in light mesons are denoted as $k_1$ and $k_2$ respectively(see
Fig.\ref{fig:transition}):
 \begin{eqnarray}
 k_1=(0,\frac{m_{B_{(s)}}}{\sqrt{2}}x_1,\textbf{k}_{1\perp})\;,\;k_2=(\frac{m_{B_{(s)}}}{\sqrt{2}}
 x_2\eta,0,\textbf{k}_{2\perp})\;.\label{eq:fmomentum}
 \end{eqnarray}

%
% %---------------------------------------
% \subsection{Wave Function of $B_{(s)}$ Mesons}
% %---------------------------------------
In the course of the PQCD calculations,   the light-cone wave
functions of the mesons are required. The B meson is  a heavy-light
system, whose light-cone matrix element can be decomposed as:
 \begin{eqnarray}
 &&\int_0^1\frac{d^4z}{(2\pi)^4}e^{ik_1\cdot z}\langle 0|b_{\beta}(0)\bar
 q_{\alpha}(z)|\bar B_{(s)}(P_{B_{(s)}})\rangle\nonumber\\
 &=&\frac{i}{\sqrt{2N_c}}\left\{(\not\! P_{B_{(s)}}+m_{B_{(s)}})\gamma_5\left[\phi_{B_{(s)}}(k_1)+\frac{\not\!
 n-\not\!
 v}{\sqrt{2}}\bar\phi_{B_{(s)}}(k_1)\right]\right\}_{\beta\alpha}\;,\label{Bwav:decompose}
 \end{eqnarray}
where $n=(1,0,\textbf{0}_T)$ and $v=(0,1,\textbf{0}_T)$ are
light-like unit vectors. There are two Lorentz structures in B meson
light-cone distribution amplitudes, and they obey the normalization
conditions:
 \begin{equation}
 \int\frac{d^4 k_1}{(2\pi)^4}\phi_{B_{(s)}}({
 k_1})=\frac{f_{B_{(s)}}}{2\sqrt{2N_c}}\;,\; \int \frac{d^4
 k_1}{(2\pi)^4}\bar{\phi}_{B_{(s)}}({ k_1})=0,\label{Bwave:normalization}
 \end{equation}
with $f_{B_{(s)}}$ as the decay constant of $B_{(s)}$ meson. In
principle, both the $\phi_{B_{(s)}}(k_1)$ and
$\bar\phi_{B_{(s)}}(k_1)$ contribute in B meson transitions.
However, the contribution of $\bar \phi_{B_{(s)}}(k_1)$ is usually
neglected, because its contribution is numerically small
\cite{Lu:2002ny}.
 So we will only keep the term with $\phi_{B_{(s)}}(k_1)$ in equation
 (\ref{Bwav:decompose}). In the momentum space the light cone matrix of B
meson can be expressed as:
 \begin{equation}
 \Phi_{B_{(s)}}=\frac{i}{\sqrt{6}}(\not \! P_{B_{(s)}} +m_{B_{(s)}})\gamma_5\phi_{B_{(s)}} (k_1). \label{Bwave:3variable}
 \end{equation}
Usually the hard part is independent of $k^+$ or/and $k^-$, so we
integrate one of them out from
$\phi_{B_{(s)}}(k^+,k^-,\textbf{k}_{\perp})$. With
 $b$ as the conjugate space coordinate of $\textbf{k}_{\perp}$, we can express
 $\phi_{B_{(s)}}(x,\textbf{k}_{\perp})$ in b-space by
 \begin{equation}
 \Phi_{{B_{(s)}},\alpha\beta}(x,b) = \frac{i}{\sqrt{2N_c}}
 \left[\not\! P_{B_{(s)}} \gamma_5 + m_{B_{(s)}}\gamma_{5} \right]_{\alpha\beta}
 \phi_{B_{(s)}}(x,b),\label{Bwave:bspace}
 \end{equation}
where $x$ is the momentum fraction of the light quark in B meson. In
this paper, we use the following expression for
$\phi_{B_{(s)}}(x,b)$:
 \begin{equation}
 \phi_{B_{(s)}}(x,b)=N_{B_{(s)}}x^2(1-x)^2\mbox{exp}\left[-\frac{m_{B_{(s)}}^2 x^2}{2\omega_b^2}-\frac{1}{2}(\omega_b
 b)^2\right], \label{Bwave:da}
 \end{equation}
with $N_{B_{(s)}}$ the normalization factor, which is determined by
equation (\ref{Bwave:normalization}). In recent years, a lot of
studies for $B^{\pm}$ and $B_d^0$ decays have been performed by the
PQCD approach. With the rich experimental data, the $\omega_b$ in
(\ref{Bwave:da}) is fixed as
$0.40\mbox{GeV}$~\cite{Keum:2000ph,direct,kphi,Lu:2002ny}. In our
calculation, we adopt $\omega_b=(0.40\pm0.05)\mbox{GeV}$ and
$f_B=(0.19\pm0.025)\rm{GeV}$ for B mesons. For $B_s$ meson, taking
the SU(3) breaking effects into consideration, we adopt
$\omega_{b}=(0.50\pm 0.05)\mbox{GeV}$\cite{Ali:2007ff} and
$f_{B_s}=(0.23\pm0.03)\rm{GeV}$.

%\cite{Keum:2000ph,direct,kphi} and in the explanation
%of $B\to \phi K^*$ polarization problem \cite{kphi,kphi-e}.
%
%%%%%%%%%%%%%%%%%%%%%%%%%%%%%%%%%%%%%%%%%%%%%%%%%%%%%%%%%%%
%\subsection{Light-cone distribution amplitudes of light vector mesons}
%%%%%%%%%%%%%%%%%%%%%%%%%%%%%%%%%%%%%%%%%%%%%%%%%%%%%%%%%%%

Decay constants of vector mesons are defined by:
\begin{equation}
\langle 0|\bar q_1\gamma_\mu
q_2|V(P_2,\epsilon)\rangle=f_Vm_V\epsilon_\mu,\;\;\; \langle 0|\bar
q_1\sigma_{\mu\nu}q_2|V(P_2,\epsilon)\rangle =if^T_V(\epsilon_\mu
P_{2\nu}-\epsilon_\nu P_{2\mu}).
\end{equation}
The longitudinal decay constants of charged vector mesons can be
extracted from the decay $\tau^- \to (\rho^-,K^{*-}) \nu_\tau$
\cite{Amsler:2008zz}. Neutral vector meson's longitudinal decay
constants can be determined by their electronic decay widths through
$V^0\to e^+e^-$ and the results are given in
Table~\ref{Table:Vdecayconstant}. Transverse decay constants are
mainly explored by QCD sum rules~\cite{Ball:2006eu}, which are also
collected in Table~\ref{Table:Vdecayconstant}.

%%%%%%%%%%%%%%%%%%%%%%%%%%%%%%%%%%%%%%%%%%%%%%%%%%%%%%%%%%
\begin{table}
\caption{Input values of the decay constants  for the vector mesons
(in MeV)}
\begin{tabular}{cccccccccc}
\hline\hline
   $f_\rho $   & $ f_\rho^T $   & $ f_\omega $ & $ f_\omega^T$
\ \ \
 & $ f_{K^*} $ & $ f_{K^*}^T $  & $f_\phi $    & $ f_\phi^T $  \\
\ \ \
   $ 209\pm 2$ & $ 165\pm 9$    & $ 195\pm 3$  & $ 151\pm 9$
\ \ \
 & $ 217\pm 5$ & $185\pm 10$    & $ 231\pm 4$  & $ 186\pm 9$\\
\hline \hline
\end{tabular}\label{Table:Vdecayconstant}
 \end{table}
%%%%%%%%%%%%%%%%%%%%%%%%%%%%%%%%%%%%%%%%%%%%%%%%%%%%%%%%%%

The vector meson polarization vectors $\epsilon$, which satisfy
$P\cdot \epsilon=0$, include one longitudinal polarization vector
$\epsilon_L$ and two transverse polarization vectors $\epsilon_T$.
The vector meson distribution amplitudes up to twist-3 are defined
by:
\begin{eqnarray}
\langle V(P_2,\epsilon^*_L)|\bar q_{2\beta}(z) q_{1\alpha}
(0)|0\rangle &=&\frac{1}{\sqrt{2N_c}}\int_0^1 dx e^{ixP_2\cdot z}
\left[m_V\not\! \epsilon^*_L \phi_V(x) +\not\! \epsilon^*_L\not\!
P_2 \phi_{V}^{t}(x) +m_V \phi_V^s(x)\right]_{\alpha\beta},
\nonumber\\
 \langle V(P_2,\epsilon^*_T)|\bar q_{2\beta}(z) q_{1\alpha}
(0)|0\rangle &=&\frac{1}{\sqrt{2N_c}}\int_0^1 dx e^{ixP_2\cdot z}
\left[ m_V\not\! \epsilon^*_T\phi_V^v(x)+ \not\!\epsilon^*_T\not\!
P_2\phi_V^T(x)\right.
\nonumber\\
& & \left.\;\;\;\;\;+m_V
i\epsilon_{\mu\nu\rho\sigma}\gamma_5\gamma^\mu\epsilon_T^{*\nu}
n^\rho v^\sigma \phi_V^a(x)\right ]_{\alpha\beta}\;, \label{spf}
\end{eqnarray}
for the longitudinal polarization and transverse polarizations,
respectively. Here  $x$ is the momentum fraction associated with the
$q_2$ quark.  $n$ is the moving direction of the vector meson and
$v$ is the opposite direction. These distribution amplitudes can be
related to the ones used in QCD sum rules by:
\begin{eqnarray}
&&\phi_{V}(x)=\frac{f_{V}}{2\sqrt{2N_c}}\phi_{||}(x),\;\;\;
\phi_{V}^t(x)=\frac{f_{V}^T}{2\sqrt{2N_c}}h_{||}^{(t)}(x),\nonumber\\
&&\phi_{V}^s(x)=\frac{f_{V}^T}{4\sqrt{2N_c}}
\frac{d}{dx}h_{||}^{(s)}(x),\hspace{3mm}
\phi_{V}^T(x)=\frac{f_{V}^T}{2\sqrt{2N_c}}\phi_{\perp}(x)
,\nonumber\\
&&\phi_{V}^v(x)=\frac{f_{V}}{2\sqrt{2N_c}}g_{\perp}^{(v)}(x),
\hspace{3mm}\phi_{V}^a(x)=\frac{f_{V}}{8\sqrt{2N_c}}
\frac{d}{dx}g_{\perp}^{(a)}(x).
\end{eqnarray}

The twist-2 distribution amplitudes can be expanded in terms of
Gegenbauer polynomials $C_n^{3/2}$ with the coefficients called
Gegenbauer moments $a_n$:
\begin{eqnarray}
 \phi_{||,\perp} (x)&=&6x (1-x)
 \left[1+\sum_{n=1}^{\infty} a_{n}^{||,\perp}C_n^{3/2} (t) \right],\label{phiV}
\end{eqnarray}
where $t=2x-1$. The Gegenbauer moments $a_{n}^{||,\perp}$ are mainly
determined by the technique of QCD sum rules. Here we quote the
recent numerical results
\cite{Braun:2004vf,Ball:2005vx,Ball:2006nr,Ball:2007rt} as
\begin{eqnarray}
 a_1^\parallel(K^*)  & = & 0.03 \pm 0.02,\;\;\;\;\;    a_1^\perp(K^*)   =0.04 \pm 0.03,\\
 a_2^\parallel(\rho) & = & a_2^\parallel(\omega)= 0.15 \pm 0.07,\;\;\;
 a_2^\perp(\rho)   =a_2^\perp(\omega)   =0.14 \pm 0.06,\\
 a_2^\parallel(K^*)  & = & 0.11 \pm 0.09,\;\;\;\;\;    a_2^\perp(K^*)   =0.10 \pm 0.08,\\
 a_2^\parallel(\phi) & = & 0.18 \pm 0.08,\;\;\;\;\;    a_2^\perp(\phi)  =0.14 \pm 0.07,
\end{eqnarray}
where the values are taken at $\mu=1$ GeV.

Using equation of motion, two-particle twist-3 distribution
amplitudes are related to the twist-2 LCDAs and the three-particle
twist-3 LCDAs. But in some $B\to VV$ decays, there exists the
so-called polarization problem. A reasonable way  has been suggested
to resolve this problem in the PQCD approach: one needs to adopt the
asymptotic LCDAs. As in Ref.~\cite{Ali:2007ff}, we use the
asymptotic forms for the twist-3 LCDAs:
\begin{eqnarray}
h_\parallel^{(t)}(x) & = & 3t^2 ,\;\;\;
h_{||}^{(s)}(x)   =  6 x(1-x),\\
%\frac{dh_{||}^{(s)}(x)}{dx} & = & 6 (1-2x),\\
g_\perp^{(a)}(x) & = & 6 x(1-x) ,\;\;\; g_\perp^{(v)}(x)  =
\frac{3}{4}\,(1+t^2).
%\frac{dg_\perp^{(a)}(x)}{dx}& = & 6\{ (1-2x) , \\ ,
\end{eqnarray}

%
%
%%%%%%%%%%%%%%%%%%%%%%%%%%%%%%%%%%%%%%%%%%%%%%%%%%%%%%%%%%%
%\subsection{Light-cone distribution amplitudes of axial-vectors}
%%%%%%%%%%%%%%%%%%%%%%%%%%%%%%%%%%%%%%%%%%%%%%%%%%%%%%%%%%

For the axial-vectors, the longitudinal and transverse decay
constants are defined by:
 \begin{eqnarray}
  \langle A(P_2,\epsilon)|\bar q_2 \gamma_\mu \gamma_5 q_1|0\rangle
   &= & if_{A}  m_{A} \,  \epsilon^{*}_\mu,\;\;\;
 \langle A(P_2,\epsilon)|
  \bar q_2 \sigma_{\mu\nu}\gamma_5 q_1
   |0\rangle
  =  f_{A}^{T}
 (\epsilon^{*}_{\mu} P_{2\nu} - \epsilon_{\nu}^{*} P_{2\mu}).
 \end{eqnarray}
In the SU(2) limit, due to G-parity invariance, the
longitudinal[transverse] decay constants vanish for the non-strange
$^1P_1$[$^3P_1$] states. This will affect the normalization for the
corresponding distribution amplitudes which will be discussed in the
following. For convenience, we take $f_{^3\! P_1}\equiv f$\,
[$f_{^1\! P_1}^T(\mu=1~{\rm GeV})\equiv f$] as the ``normalization
constant". The decay constants of axial vector mesons shown in
table~\ref{Table:Adecayconstant} are taken from Ref.
\cite{Yang:2005gk,Yang:2007zt}.

%%%%%%%%%%%%%%%%%%%%%%%%%%%%%%%%%%%%%%%%%%%%%%%%%%%%%%%%%%
\begin{table}
\caption{Input values of the decay constants (absolute values) for
the axial-vector mesons (in MeV). The transverse decay constants for
$^1P_1$ are evaluated at $\mu =1$ GeV.}
\begin{tabular}{cccccccccc}
\hline\hline
   $f_{a_1(1260)} $   & $ f_{ f_1 (1^3P_1)} $   & $ f_{ f_8 (1^3P_1)} $ & $ f_{ K_{1A}}$
\ \ \
  & $ f^T_{b_1(1235)} $ & $ f^T_{h_1 (1^1P_1)} $  & $f^T_{h_8 (1^1P_1)} $    & $ f^T_{K_{1B}} $  \\
\ \ \
   $ 238\pm 10$ & $ 245\pm 13$    & $ 239\pm 13$  & $ 250\pm 13$
\ \ \
 & $  180\pm 8$ & $180\pm 12$    & $ 190\pm 10$  & $ 190\pm 10$\\
\hline \hline
\end{tabular}\label{Table:Adecayconstant}
 \end{table}
%%%%%%%%%%%%%%%%%%%%%%
%%%%%%%%%%%%%%%%%%%%%%%%%%%%%%%%%%%%%%%%%%%%%%%%%%%%%%%%%%

Distribution amplitudes for axial-vectors with quantum numbers
$J^{PC}=1^{++}$ or $1^{+-}$ are defined by:
\begin{eqnarray}
\langle A(P_2,\epsilon^*_L)|\bar q_{2\beta}(z) q_{1\alpha}
(0)|0\rangle &=&\frac{-i}{\sqrt{2N_c}}\int_0^1 dx e^{ixp\cdot z}
\left[-m_A\gamma_5\not\! \epsilon^*_L \phi_A(x) -\not\!
\epsilon^*_L\not\! P_2 \gamma_5\phi_{A}^{t}(x) -m_A\gamma_5
\phi_A^s(x)\right]_{\alpha\beta},
\nonumber\\
 \langle A(P_2,\epsilon^*_T)|\bar q_{2\beta}(z) q_{1\alpha}
(0)|0\rangle &=&\frac{-i}{\sqrt{2N_c}}\int_0^1 dx e^{ixp\cdot z}
\left[-m_A\gamma_5\not\! \epsilon^*_T\phi_A^v(x) -
\not\!\epsilon^*_T\not\! P_2\gamma_5\phi_A^T(x)\right.
\nonumber\\
& & \left.\;\;\;\;\;-m_A
i\epsilon_{\mu\nu\rho\sigma}\gamma^\mu\epsilon_T^{*\nu} n^\rho
v^\sigma \phi_A^a(x)\right ]_{\alpha\beta}\;.
\label{eq:transverseWVA}
\end{eqnarray}
Besides the factor $i\gamma_5$ from the left hand, axial-vector
mesons' distribution amplitudes can be related to the vector ones by
making the following replacement:
\begin{eqnarray}
\phi_V\to \phi_A,\;\;\phi_V^t\to  \phi_A^t,\;\;\phi_V^s\to
 \phi_A^s,\nonumber\\
\phi_V^T\to  \phi_A^T,\;\;\phi_V^v\to \phi_A^v,\;\;\phi_V^a\to
\phi_A^a.\label{eq:relationofAVDA}
\end{eqnarray}
These distribution amplitudes can be related to the ones
calculated in QCD sum rules by:
\begin{eqnarray}
 &&\phi_{A}(x)=\frac{f}{2\sqrt{2N_c}}\phi_{||}(x), \;\;\;\phi_{A}^t(x)=\frac{f}{2\sqrt{2N_c}}h_{||}^{(t)}(x),\nonumber\\
&&\hspace{-5mm}\phi_{A}^s(x)=\frac{f}{4\sqrt{2N_c}}
\frac{d}{dx}h_{\parallel}^{(s)}(x),\hspace{3mm}
\phi_{A}^T(x)=\frac{f}{2\sqrt{2N_c}}\phi_{\perp}(x)
,\nonumber\\
&&\hspace{-5mm}\phi_{A}^v(x)=\frac{f}{2\sqrt{2N_c}}g_{\perp}^{(v)}(x),
\hspace{3mm}\phi_{A}^a(x)=\frac{f}{8\sqrt{2N_c}}
\frac{d}{dx}g_{\perp}^{(a)}(x),
\end{eqnarray}
where we use $f$ as the ``normalization" constant for both
longitudinally and transversely polarized mesons.

In the isospin limit,  $\phi_\parallel$, $g_\perp^{(a)}$ and
$g_\perp^{(v)}$ are symmetric [antisymmetric] under the replacement
$x\leftrightarrow 1-x$, while $\phi_\perp$, $h_{||}^{(t)}$, and
$h_{||}^{(s)}$ are antisymmetric [symmetric] for non-strange
$1^3P_1$ [$1^1P_1$] states. In the above, we have taken $f_{^3\!
P_1}^T=f_{^3\! P_1}=f$\, [$f_{^1\! P_1}=f_{^1\! P_1}^T(\mu=1~{\rm
GeV})=f$], thus we have
 \begin{eqnarray}
 \langle 1^3P_1(P,\epsilon)|
  \bar q_1 \sigma_{\mu\nu}\gamma_5 q_2
   |0\rangle
  &= & f_{^3P_1}^{T} a_0^{\perp,^3P_1} \,
(\epsilon^{*}_{\mu} P_{\nu} - \epsilon_{\nu}^{*} P_{\mu}),
 \\
  \langle 1^1P_1(P,\epsilon)|\bar q_1 \gamma_\mu \gamma_5 q_2|0\rangle
   &= & if_{^1P_1} a_0^{\parallel,^1P_1} \, m_{^1P_1} \,  \epsilon^{*}_\mu,
 \end{eqnarray}
where $a_0^{\perp,^3P_1}$ and  $a_0^{\parallel,^1P_1}$ are the
Gegenbauer zeroth moments. Then the normalization conditions of the
distribution amplitudes are given by
\begin{eqnarray}
 \int_0^1 dx \phi_\perp (x) &=& a^\perp_0\label{eq:normaizlationofA1}
 \end{eqnarray}
for $1^3P_1$ states and
 \begin{eqnarray}
  \int_0^1 dx \phi_{||}(x)=a^{||}_0\label{eq:normaizlationofA2}
  \end{eqnarray}
for $1^1P_1$ states. The zeroth Gegenbauer moments
$a_0^{\perp,^3P_1}$ and $a_0^{\parallel,^1P_1}$, characterizing the
breaking of flavor SU$(3)$ symmetry, are non-zero for only strange
mesons. We normalize the distribution amplitude
$\phi_\parallel\big[\phi_\perp\big]$ of the $1^3P_1\big[1^1P_1\big]$
states as
 \begin{equation}
 \int_0^1 dx\phi_{||}(x)=1\bigg[\int_0^1 dx\phi_{\perp}(x)=1\bigg].
 \end{equation}
For convenience, we formally define $a_0^{||}=1$ for the $1^3P_1$
states so that we can use Eq.~(\ref{eq:normaizlationofA2}) as the
normalization condition. Similarly, we also define $a_0^{\perp}=1$
for $1^1P_1$ states so that $\phi_\perp (x)$ has a correct
normalization.

%Thus, in order to make (\ref{eq:normaizlationofA1}) and
%(\ref{eq:normaizlationofA2}) available for both $1^3P_1$ and
%$1^1P_1$ states, $a_0^{||,1^3P_1}=1$ and $a_0^{\perp,1^1P_1}=1$ are
%defined formally.

%For convenience, we formally define $a_0^{||}=1$ for the $1^3P_1$
%states so that we can use Eq.~(\ref{eq:normaizlationofA}) as the
%normalization condition. Similarly, we also define $a_0^{\perp}=1$
%for $1^1P_1$ states so that $\phi_\perp (x)$ has a correct
%normalization.

%
%%%%%%%%%%%%%%%%%%%%%%%%%%%%%%%%%%%%%%%%%%%%%%%%%%%%%%%%%%%%%%%%%%%%%%%%%%%%%%%
\begin{table}[ttb]
\caption{Gegenbauer moments of $\phi_\perp$ and $\phi_{||}$ for
$1^3P_1$ and $1^1P_1$ mesons evaluated in Ref. \cite{Yang:2007zt},
where the values are taken at $\mu=1$ GeV.}
\label{tab:AxialGegenbauer} $
\begin{array}{|c|c|c|c|c|c|}
 \hline
  a_2^{||, a_1(1260)}       & a_2^{||,f_1^{^3P_1}}        & a_2^{||,f_8^{^3P_1}}         & a_2^{||, K_{1A}}
                            & \multicolumn{2}{|c|}{a_1^{||, K_{1A}}}\\
  \hline
 -0.02\pm 0.02              &  -0.04\pm 0.03 & -0.07\pm 0.04 & -0.05\pm 0.03
                            & \multicolumn{2}{c|}{ {0.00\pm 0.26}}\\
 \hline
  a_1^{\perp, a_1(1260)}    & a_1^{\perp,f_1^{^3P_1}}     & a_1^{\perp,f_8^{^3P_1}}      & a_1^{\perp, K_{1A}}
                            & a_0^{\perp, K_{1A}}         & a_2^{\perp, K_{1A}}\\
  \hline
 -1.04\pm 0.34              & -1.06\pm 0.36               & -1.11\pm 0.31                & -1.08\pm 0.48
                            &  0.08\pm 0.09               & 0.02\pm 0.20\\
 \hline
 a_1^{||, b_1(1235)}        & a_1^{||,h_1^{^1P_1}}        &a_1^{||,h_8^{^1P_1}}          & a_1^{||, K_{1B}}
                            & a_0^{||, K_{1B}}            & a_2^{||, K_{1B}} \\
 \hline
    -1.95\pm 0.35           & -2.00\pm 0.35               & -1.95\pm 0.35                & -1.95\pm 0.45
                            &  0.14\pm 0.15               &  0.02\pm 0.10\\
 \hline
  a_2^{\perp, b_1(1235)}    & a_2^{\perp,h_1^{^1P_1}}     & a_2^{\perp,h_8^{^1P_1}}      & a_2^{\perp, K_{1B}}
                            & \multicolumn{2}{|c|}{a_1^{\perp, K_{1B}}} \\
 \hline
            0.03\pm 0.19    &  0.18\pm 0.22               &   0.14\pm 0.22               & -0.02\pm 0.22
                            &\multicolumn{2}{c|}{0.17\pm 0.22}\\
 \hline
\end{array}
$
\end{table}
%%%%%%%%%%%%%%%%%%%%%%%%%%%%%%%%%%%%%%%%%%%%%%%%%%%%%%%%%%%%%%%%%%%%%%%%%%%%%%%

Up to conformal spin 6, twist-2 distribution amplitudes for
axial-vector mesons can be expanded as:
\begin{eqnarray}
 \phi_\parallel(x) & = & 6 x \bar x \left[ a_0^\parallel + 3
a_1^\parallel\, t +
a_2^\parallel\, \frac{3}{2} ( 5t^2  - 1 ) \right], \label{eq:lcda-1p1-t2-1}\\
 \phi_\perp(x) & = & 6 x \bar x \left[ a_0^\perp + 3 a_1^\perp\, t +
a_2^\perp\, \frac{3}{2} ( 5t^2  - 1 ) \right],
\label{eq:lcda-3p1-t2-2}
\end{eqnarray}
where the Gegenbauer moments are calculated in Refs.
\cite{Yang:2005gk,Yang:2007zt} shown in table
\ref{tab:AxialGegenbauer}. From the results in table
\ref{tab:AxialGegenbauer}, we can see that there are large
uncertainties in Gegenbauer moments which can inevitably induce
large uncertainties to form factors and branching ratios. We hope
the uncertainties could be reduced in future studies in order to
make more precise predictions.

As for twist-3 LCDAs, we use the following form:
\begin{eqnarray}
 g_\perp^{(v)}(x) & = & \frac{3}{4} a_0^\parallel (1+t^2)
 + \frac{3}{2}\, a_1^\parallel\, t^3,\;\;\;
 g_\perp^{(a)}(x)  =  6 x \bar x ( a_0^\parallel + a_1^\parallel t),
 \\
 h_\parallel^{(t)}(x) &= & 3a_0^\perp t^2+ \frac{3}{2}\,a_1^\perp\,t (3 t^2-1)
 ,\;\;\;
 h_\parallel^{(s)}(x)  =  6x\bar x( a_0^\perp + a_1^\perp t ).
 \end{eqnarray}

In the following analysis, we will use $a_1$ to denote $a_1(1260)$,
$b_1$ to denote $b_1(1235)$ for simplicity. It is also similar for
$K_1$ and $f_1,h_1$.

%%%%%%%%%%%%%%%%%%%%%%%%%%%%%%%%%%%%%%%%%%%%%%%%%%%%%%%%%%
\section{$B\to V$, $B\to A$  form factors and semileptonic $B\to Al\bar\nu$ decays}\label{section:formfactor}
%%%%%%%%%%%%%%%%%%%%%%%%%%%%%%%%%%%%%%%%%%%%%%%%%%%%%%%%%%

\subsection{PQCD approach}

%%%%%%%%%%%%%%%%%%%%%%%%%%%%%%%%%%%%%%%%%%%%%%%%
\begin{figure}[tb]
\begin{center}
\psfig{file=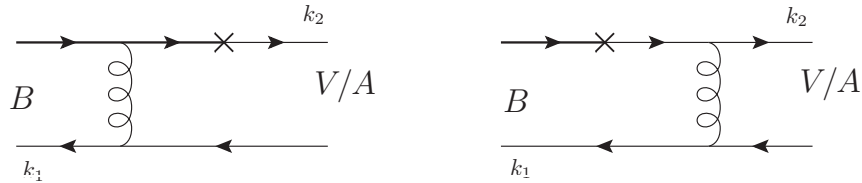,width=12.0cm,angle=0}
\end{center}
\caption{Feynman diagrams for transition of B meson to a vector or
axial vector meson. The crosses represent Lorentz structures of the
currents.}\label{fig:transition}
\end{figure}
%%%%%%%%%%%%%%%%%%%%%%%%%%%%%%%%%%%%%%%%%%%%%%%%%%%%%

The basic idea of the PQCD approach is that it takes into account
the intrinsic transverse momentum of valence quarks. The decay
amplitude, taking the first diagram in Fig.~\ref{fig:transition} as
an example, can be expressed as a convolution of wave functions
$\phi_B$, $\phi_2$ and hard scattering kernel $T_H$ with both
longitudinal and transverse momenta:
\begin{eqnarray}
{\cal M}=\int^1_0dx_1dx_2\int\frac{
d^2{\vec{k}}_{1T}}{(2\pi)^2}\frac{d^2{\vec{k}}_{2T}}{(2\pi)^2}\phi_B(x_1,{\vec{k}}_{1T},P_B,t)
T_H(x_1,x_2,{\vec{k}}_{1T},{\vec{k}}_{2T},t)
\phi_2(x_2,{\vec{k}}_{2T},P_2,t).
\end{eqnarray}
Usually it is convenient to compute the amplitude in coordinate
space. Through Fourier transformation, the above equation can be
expressed by:
\begin{eqnarray}
{\cal M}=\int^1_0dx_1dx_2\int
{d^2{\vec{b}}_{1}}{d^2{\vec{b}}_{2}}{\phi}_B(x_1,{\vec{b}}_{1},P_B,t)
T_H(x_1,x_2,{\vec{b}}_{1},{\vec{b}}_{2},t){\phi}_2(x_2,{\vec{b}}_{2},P_2,t).
\end{eqnarray}

\begin{figure}%[]
\begin{center}
\includegraphics[width=3.5cm]{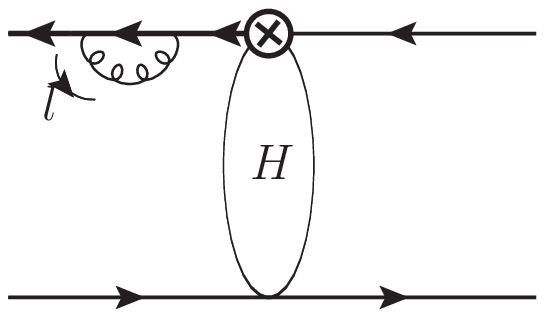}
\hspace{3mm}
\includegraphics[width=3.5cm]{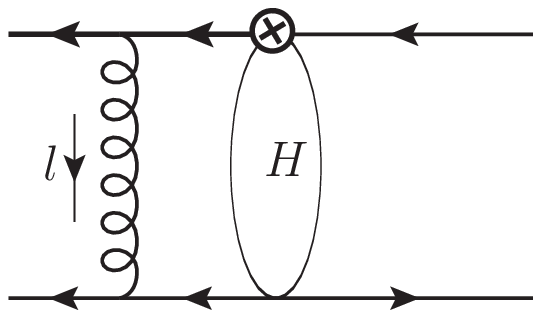}
\vspace{5mm}
\includegraphics[width=3.5cm]{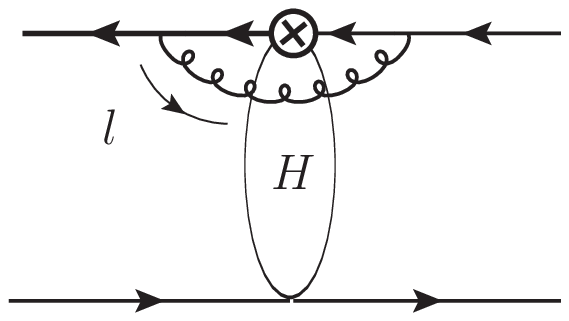}
\hspace{3mm}
\includegraphics[width=3.5cm]{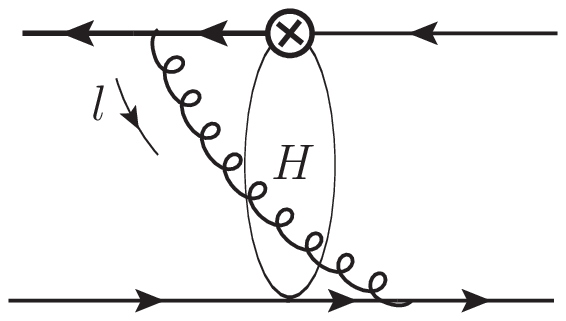}
\end{center}
\vspace{-.7cm} \caption{${O(\alpha_{s})}$ corrections to the hard
scattering kernel ${H}$.} \label{Feyn:alpha}
\end{figure}

This derivation is mainly concentrated on tree level diagrams, but
actually we have to take into account some loop effects which can
give sizable corrections. The ${\cal O}(\alpha_s)$ radiative
corrections to hard scattering process ${H}$ are depicted in
Fig.~\ref{Feyn:alpha}. In general, individual higher order diagrams
may suffer from two types of infrared divergences: soft and
collinear. Soft divergence comes from the region of a loop momentum
where all it's momentum components vanish:
\begin{eqnarray}
l^{\mu}=(l^+,l^-,\vec{l}_T)=(\Lambda,\Lambda,\vec{\Lambda}),
\end{eqnarray}
where $\Lambda$ is the typical scale for hadronization. Collinear
divergence originates from the gluon momentum region which is
parallel to the massless quark momentum,
\begin{eqnarray}
l^{\mu}=(l^+,l^-,\vec{l}_T) \sim
(m_B,{\Lambda}^2/m_B,\vec{\Lambda}).
\end{eqnarray}
In both cases, the loop integration corresponds to ${\int d^4 l/l^4
\sim \log{\Lambda}}$, thus logarithmic divergences are generated. It
has been shown order by order in perturbation theory that these
divergences can be separated from the hard kernel and absorbed into
meson wave functions using eikonal approximation \cite{Li:1994iu}.
But when soft and collinear momentum overlap, there will be double
logarithm divergences in the first two diagrams of
Fig.~\ref{Feyn:alpha}. These large double logarithm can be resummed
into the Sudakov factor whose explicit form is given in Appendix
\ref{PQCDfunctions}.

Furthermore, there are also another type of double logarithm which
comes from the loop correction for the weak decay vertex correction.
The left diagram in Fig.~\ref{fig:transition}  gives an amplitude
proportional to $1/((1-x_2)^2 x_1)$. In the threshold region with
$(1-x_2)\to 0$  [{(to be precise, $(1-x_2)\sim
O(\Lambda_{QCD}/m_B)$)}], additional soft divergences are associated
with the internal quark at higher orders. The QCD loop corrections
to the electro-weak vertex can produce the double logarithm
$\alpha_s\ln^2 (1-x_2)$ and resummation of this type of double
logarithms lead to the Sudakov factor $S_t(x_2)$. Similarly,
resummation of $\alpha_s\ln^2 x_1$ due to loop corrections in the
other diagram leads to the Sudakov factor $S_t(x_1)$. These double
logarithm can also be factored out from the hard part and grouped
into the quark jet function. Resummation of the double logarithms
results in the threshold factor \cite{Li:2001ay}. This factor
decreases faster than any other power of ${x}$ as ${x\rightarrow
0}$, which modifies the behavior in the endpoint region to make pQCD
approach more self-consistent. For simplicity, this factor has been
parameterized in a form which is independent on channels, twists and
flavors \cite{Li:2002mi}.

Combing all the elements together, we can get the typical
factorization formulae in the PQCD approach:
\begin{eqnarray}
{\cal M}&=&\int^1_0dx_1dx_2\int
{d^2{\vec{b}}_{1}}{d^2{\vec{b}}_{2}}{(2\pi)^2}{\phi}_B(x_1,{\vec{b}}_{1},P_B,t)\nonumber\\
&&\times
T_H(x_1,x_2,Q,{\vec{b}}_{1},{\vec{b}}_{2},t){\phi}_2(x_2,{\vec{b}}_{2},P_2,t)
S_t(x_2)\exp[-S_B(t)-S_2(t)].
\end{eqnarray}

\subsection{$B\to V$ form factors}

$\bar B\to V$ form factors are defined under the conventional form
as follows:
 \begin{eqnarray}
  \langle V(P_2,\epsilon^*)|\bar q\gamma^{\mu}b|\bar B(P_B)\rangle
   &=&-\frac{2V(q^2)}{m_B+m_V}\epsilon^{\mu\nu\rho\sigma}
     \epsilon^*_{\nu}P_{B\rho}P_{2\sigma}, \nonumber\\
  \langle V(P_2,\epsilon^*)|\bar q\gamma^{\mu}\gamma_5 b|\bar
  B(P_B)\rangle
   &=&2im_V A_0(q^2)\frac{\epsilon^*\cdot q}{q^2}q^{\mu}
    +i(m_B+m_V)A_1(q^2)\left[\epsilon^*_{\mu}
    -\frac{\epsilon^*\cdot q}{q^2}q^{\mu} \right] \nonumber\\
    &&-iA_2(q^2)\frac{\epsilon^*\cdot q}{m_B+m_V}
     \left[ (P_B+P_2)^{\mu}-\frac{m_B^2-m_V^2}{q^2}q^{\mu} \right],\nonumber\\
  \langle V(P_2,\epsilon^*)|\bar q\sigma^{\mu\nu}q_{\nu}b|\bar
  B(P_B)\rangle
   &=&-2iT_1(q^2)\epsilon^{\mu\nu\rho\sigma}
     \epsilon^*_{\nu}P_{B\rho}P_{2\sigma}, \nonumber\\
  \langle V(P_2,\epsilon^*)|\bar q\sigma^{\mu\nu}\gamma_5q_{\nu}b|\bar
  B(P_B)\rangle
   &=&T_2(q^2)\left[(m_B^2-m_V^2)\epsilon^{*\mu}
       -(\epsilon^*\cdot q)(P_B+P_2)^{\mu} \right]\nonumber\\
   &&+T_3(q^2)(\epsilon^*\cdot q)\left[
       q^{\mu}-\frac{q^2}{m_B^2-m_V^2}(P_B+P_2)^{\mu}\right],\label{eq:BtoVformfactors}
 \end{eqnarray}
where $q=P_B-P_2$, and the relation
$2m_VA_0(0)=(m_B+m_V)A_1(0)-(m_B-m_V)A_2(0)$ is obtained in order to
cancel the pole at $q^2=0$.

The factorization formulae are given as:
 \begin{eqnarray}
 V(q^2)&=&8 \pi C_F m_B^2 (1+r_2) \int_0^1
        dx_1dx_2\int_0^{\infty}b_1db_1b_2db_2 \phi_B(x_1,b_1)\nonumber\\
        &&\times\left\{ \bigg[\phi_V^T(x_2)-r_2\big((1-x_2)(\phi_V^v(x_2)+\phi_V^a(x_2))
        +\frac{2}{\eta}\phi_V^a(x_2)\big)\bigg]\right.\nonumber\\
        &&\left.\times h_e(x_1,(1-x_2)\eta,b_1,b_2)\alpha_s(t_e^1)
        \mbox{exp}[-S_{ab}(t_e^1)]S_t(x_2)\right.\nonumber\\
        &&\left.- r_2\big(\phi_V^a(x_2)-\phi_V^v(x_2)\big)
        h_e(1-x_2,x_1\eta,b_2,b_1)\alpha_s(t_e^2)\mbox{exp}[-S_{ab}(t_e^2)]S_t(x_1)\right\},
 \end{eqnarray}
 \begin{eqnarray}
 A_0(q^2)&=&8 \pi C_F m_B^2 \int_0^1
        dx_1dx_2\int_0^{\infty}b_1db_1b_2db_2 \phi_B(x_1,b_1)\nonumber\\
        &&\times\left\{\bigg[-\phi_V(x_2)(x_2\eta-\eta-1)-r_2\bigg((-3+\frac{2}{\eta}+2x_2)\phi_V^s(x_2)
        +(1-2x_2)\phi_V^t(x_2)\bigg)\bigg]\right.\nonumber\\
        &&\left.\times h_e(x_1,(1-x_2)\eta,b_1,b_2)\alpha_s(t_e^1)
        \mbox{exp}[-S_{ab}(t_e^1)]S_t(x_2)\right.\nonumber\\
        &&\left.-2r_2\phi_V^s(x_2)
        h_e(1-x_2,x_1\eta,b_2,b_1)\alpha_s(t_e^2)\mbox{exp}[-S_{ab}(t_e^2)]S_t(x_1)\right\},
 \end{eqnarray}
 \begin{eqnarray}
 A_1(q^2)&=&8 \pi C_F m_B^2 (1-r_2)\int_0^1
        dx_1dx_2\int_0^{\infty}b_1db_1b_2db_2 \phi_B(x_1,b_1)\nonumber\\
        &&\times\left\{ \bigg[\eta\phi_V^T(x_2)-r_2\bigg((\phi_A^a(x_2)
        +\phi_V^v(x_2))(x_2-1)\eta-2\phi_V^v(x_2)\bigg)\bigg]\right.\nonumber\\
        &&\left.\times h_e(x_1,(1-x_2)\eta,b_1,b_2)\alpha_s(t_e^1)
        \mbox{exp}[-S_{ab}(t_e^1)]S_t(x_2)\right.\nonumber\\
        &&\left.-r_2\eta(\phi_V^a(x_2)-\phi_V^v(x_2))
        h_e(1-x_2,x_1\eta,b_2,b_1)\alpha_s(t_e^2)\mbox{exp}[-S_{ab}(t_e^2)]S_t(x_1)\right\},
 \end{eqnarray}
 \begin{eqnarray}
 T_1(q^2)&=&8 \pi C_F m_B^2 \int_0^1
        dx_1dx_2\int_0^{\infty}b_1db_1b_2db_2 \phi_B(x_1,b_1)\nonumber\\
        &&\times\left\{\bigg[\phi_V^T(x_2)(1+\eta-x_2\eta)-r_2\bigg((-3
        +\frac{2}{\eta}+2x_2)\phi_V^a(x_2)+(1-2x_2)\phi_V^v(x_2)\bigg)\bigg]\right.\nonumber\\
        &&\left.\times h_e(x_1,(1-x_2)\eta,b_1,b_2)\alpha_s(t_e^1)
        \mbox{exp}[-S_{ab}(t_e^1)]S_t(x_2)\right.\nonumber\\
        &&\left.-r_2(\phi_V^a(x_2)-\phi_V^v(x_2))
        h_e(1-x_2,x_1\eta,b_2,b_1)\alpha_s(t_e^2)\mbox{exp}[-S_{ab}(t_e^2)]S_t(x_1)\right\},
 \end{eqnarray}
 \begin{eqnarray}
 T_2(q^2)&=&8 \pi C_F m_B^2 \int_0^1
        dx_1dx_2\int_0^{\infty}b_1db_1b_2db_2 \phi_B(x_1,b_1)\nonumber\\
        &&\times\left\{\bigg[\phi_V^T(x_2)\eta(1+\eta-x_2\eta)
        +r_2\eta\bigg((-3+\frac{2}{\eta}+2x_2)\phi_V^v(x_2)+(1-2x_2)\phi_V^a(x_2)\bigg)\bigg]\right.\nonumber\\
        &&\left.\times h_e(x_1,(1-x_2)\eta,b_1,b_2)\alpha_s(t_e^1)
        \mbox{exp}[-S_{ab}(t_e^1)]S_t(x_2)\right.\nonumber\\
        &&\left.-r_2\eta(\phi_V^a(x_2)-\phi_V^v(x_2))
        h_e(1-x_2,x_1\eta,b_2,b_1)\alpha_s(t_e^2)\mbox{exp}[-S_{ab}(t_e^2)]S_t(x_1)\right\},
 \end{eqnarray}
 \begin{eqnarray}
 T_3(q^2)&=&\frac{T_2(\eta)}{\eta}-\frac{1}{\eta} 16r_2\pi C_F m_B^2 \int_0^1
        dx_1dx_2\int_0^{\infty}b_1db_1b_2db_2 \phi_B(x_1,b_1)\nonumber\\
        &&\times\phi_V(x_2) h_e(x_1,(1-x_2)\eta,b_1,b_2)\alpha_s(t_e^1)
        \mbox{exp}[-S_{ab}(t_e^1)]S_t(x_2).
 \end{eqnarray}
With terms suppressed by $r_2^2$ neglected, $V_2(q^2)$ can be
expressed linearly by $V_0(q^2)$ and $V_1(q^2)$:
 \begin{eqnarray}
 A_2(q^2)&=&\frac{1}{\eta}\big[(1-r_2)^2A_1(q^2)-2r_2(1-r_2)A_0(q^2)\big].%\nonumber\\
%        &&+32(1-r_2)r_2^2\frac{\eta-1}{\eta^2}\pi C_F m_B^2 \int_0^1
%        dx_1dx_2\int_0^{\infty}b_1db_1b_2db_2 \phi_B(x_1,b_1)\nonumber\\
%        &&\times(\phi_A^s(x_2)+\phi_A^t(x_2)) h_e(x_1,(1-x_2)\eta,b_1,b_2)\alpha_s(t_e^1)
%        \mbox{exp}[-S_{ab}(t_e^1)]S_t(x_2)
 \end{eqnarray}
The definitions of the function $S_{ab}(t)$ in Sudakov exponent
${\rm{exp}}[-S_{ab}(t)]$, the factorization scales $t_e^i$s and hard
functions $h_e$, are given in Appendix~\ref{PQCDfunctions}.

%%%%%%%%%%%%%%%%%%%%%%%%%%%%%%%%%%%%%%%%%%%%%%%%%%%%%%%%%%
\begin{table}
\caption{$B\to V$ form factors at maximally recoil, i.e. $q^2=0$.
The first error comes from decay constants and shape parameters
$\omega_b$ of $B$ mesons; while the second one is from the hard
scale $t_e$, the threshold resummation parameter $c$  and
$\Lambda_{QCD}$.  }
\begin{center}
\begin{tabular}{cc|c|c|c|c|c}
\hline \hline
         &       & $B\to\rho$ &$B\to K^*$     &$B\to\omega$ &$B_s\to K^*$  &$B_s\to\phi$ \\
 \hline
LFQM\cite{Cheng:2003sm}&$V$ &$0.27$      &$ 0.31$        &             &              &        \\
      & $A_0$  &$0.28$      &$0.31$         &             &              &        \\
     & $A_1$  &$0.22$      &$0.26$         &             &              &        \\
      & $A_2$  &$0.20$      &$ 0.24$        &             &              &        \\
\hline
LCSR\cite{Ball:2004rg}&$V$&$0.323$  &$0.411$        &$0.293$      &$0.311$       &$0.434$ \\
      & $A_0$  &$0.303$     &$0.374$        &$0.281$      &$0.360$       &$0.474$ \\
      & $A_1$  &$0.242$     &$0.292$        & $0.219$     &$0.233$       &$0.311$ \\
      & $A_2$  &$0.221$     &$0.259$        &$0.198$      &$0.181$       &$0.234$ \\
      & $T_2$  &$0.267$     &$0.333$        &$0.242$      &$0.260$       &$0.349$ \\
\hline
LQCD\cite{DelDebbio:1997kr} &$V$&$0.35 $     &               &             &              &        \\
     & $A_0$  &$0.30 $     &               &             &              &        \\
     & $A_1$  &$0.27$      &               &             &              &        \\
      & $A_2$  &$0.26$      &               &             &              &        \\
  \cite{Becirevic:2006nm}  & $T_1$  &         &$0.24$  &             &              &        \\
\hline SCET LCQM\cite{Lu:2007sg}
     & $V$    &$0.298$     &$0.339$        &$0.275$      &$0.323$       &$0.329$ \\
     & $A_0$  &$0.260$     &$0.283$        &$0.240$      &$0.279$       &$0.279$ \\
     & $A_1$  &$0.227$     &$0.248$        &$0.209$      &$0.228$       &$0.232$ \\
      & $A_2$  &$0.215$     &$0.233$        &$0.198$      &$0.204$       &$0.210$ \\
     & $T_1=T_2$  &$0.260$     &$0.290$        &$0.239$      &$0.271$       &$0.276$ \\
     & $T_3$  &$0.184$     &$0.194$        &$0.168$      &$0.165$       &$0.170$\\
\hline
This work  & $V$    &$0.21_{-0.04-0.02}^{+0.05+0.03}$     &$0.25_{-0.05-0.02}^{+0.06+0.04}$        &$0.19_{-0.04-0.02}^{+0.04+0.03}$      &$0.20_{-0.04-0.02}^{+0.04+0.03}$   &$0.26_{-0.05-0.03}^{+0.05+0.04}$\\
       & $A_0$  &$0.25_{-0.05-0.03}^{+0.06+0.04}$     &$0.31_{-0.06-0.03}^{+0.07+0.05}$       &$0.23_{-0.04-0.02}^{+0.05+0.03}$      &$0.24_{-0.04-0.02}^{+0.05+0.04}$   &$0.31_{-0.06-0.03}^{+0.06+0.05}$\\
       & $A_1$  &$0.16_{-0.03-0.02}^{+0.04+0.02}$     &$0.19_{-0.04-0.02}^{+0.04+0.03}$       &$0.15_{-0.03-0.01}^{+0.03+0.02}$      &$0.15_{-0.03-0.01}^{+0.03+0.02}$   &$0.18_{-0.03-0.02}^{+0.04+0.03}$\\
      & $A_2$  &$0.13_{-0.03-0.01}^{+0.03+0.02}$     &$0.14_{-0.03-0.01}^{+0.03+0.02}$       &$0.12_{-0.02-0.01}^{+0.03+0.02}$      &$0.11_{-0.02-0.01}^{+0.02+0.01}$   &$0.12_{-0.02-0.01}^{+0.03+0.01}$\\
      & $T_1=T_2$  &$0.19_{-0.04-0.02}^{+0.04+0.03}$     &$0.23_{-0.05-0.02}^{+0.05+0.03}$       &$0.18_{-0.04-0.02}^{+0.04+0.02}$      &$0.18_{-0.03-0.02}^{+0.04+0.02}$   &$0.23_{-0.04-0.02}^{+0.05+0.03}$\\
      & $T_3$  &$0.17_{-0.03-0.02}^{+0.04+0.02}$     &$0.20_{-0.04-0.02}^{+0.05+0.03}$       &$0.15_{-0.03-0.02}^{+0.04+0.02}$      &$0.16_{-0.03-0.02}^{+0.03+0.02}$   &$0.19_{-0.04-0.02}^{+0.04+0.03}$\\
% \hline
% Asymptotic  & $V$    &$0.21_{-0.04-0.00}^{+0.05+0.00}$     &$0.25_{-0.05-0.00}^{+0.05+0.00}$        &$0.20_{-0.04-0.00}^{+0.04+0.00}$      &$0.21_{-0.03-0.01}^{+0.04+0.00}$   &$0.25_{-0.04-0.01}^{+0.05+0.00}$\\
%       & $A_0$  &$0.25_{-0.05-0.01}^{+0.05+0.00}$     &$0.30_{-0.05-0.01}^{+0.06+0.00}$       &$0.24_{-0.04-0.01}^{+0.05+0.00}$      &$0.26_{-0.04-0.01}^{+0.05+0.00}$   &$0.30_{-0.05-0.01}^{+0.05+0.00}$\\
%       & $A_1$  &$0.17_{-0.03-0.00}^{+0.04+0.00}$     &$0.19_{-0.03-0.00}^{+0.04+0.00}$       &$0.15_{-0.03-0.00}^{+0.03+0.00}$      &$0.16_{-0.02-0.01}^{+0.03+0.00}$   &$0.18_{-0.03-0.01}^{+0.03+0.00}$\\
%%       & $A_2$  &$0.13_{-0.02-0.00}^{+0.03+0.00}$     &$0.14_{-0.03-0.00}^{+0.03+0.00}$       &$0.13_{-0.02-0.00}^{+0.03+0.00}$      &$0.12_{-0.02-0.01}^{+0.02+0.00}$   &$0.13_{-0.02-0.01}^{+0.02+0.00}$\\
%      & $T_1=T_2$  &$0.20_{-0.04-0.00}^{+0.04+0.00}$     &$0.23_{-0.04-0.00}^{+0.05+0.00}$       &$0.18_{-0.03-0.00}^{+0.04+0.00}$      &$0.19_{-0.03-0.01}^{+0.04+0.00}$   &$0.22_{-0.04-0.01}^{+0.04+0.00}$\\
%      & $T_3$  &$0.13_{-0.02-0.00}^{+0.03+0.00}$     &$0.13_{-0.02-0.00}^{+0.03+0.00}$       &$0.12_{-0.02-0.00}^{+0.03+0.00}$      &$0.11_{-0.02-0.01}^{+0.02+0.00}$   &$0.12_{-0.02-0.01}^{+0.02+0.00}$\\
 \hline \hline
\end{tabular}\label{Tab:BtoVformfactor}
\end{center}
\end{table}
%%%%%%%%%%%%%%%%%%%%%%%%%%%%%%%%%%%%%%%%%%%%%%%%%%%%%%%%%%

The numerical results for the form factors at maximally recoil point
are collected in table~\ref{Tab:BtoVformfactor}. The first error
comes from decay constants and shape parameter $\omega_b$ of
$B_{(s)}$ meson; while the second one is from hard scales $t_e^i$s,
the threshold resummation parameter $c=0.4\pm0.1$ and
$\Lambda_{QCD}\big((0.25\pm0.05)\rm{GeV}\big)$. To make a
comparison, we also collect the results using other approaches
\cite{Cheng:2003sm,Ball:2004rg,DelDebbio:1997kr,Becirevic:2006nm,Lu:2007sg}.
From table~\ref{Tab:BtoVformfactor}, we can see that most of our
results are consistent with others within theoretical errors.

\subsection{$B\to A$ form factors}
Following Ref. \cite{Cheng:2004yj},  the $\bar B\to A$ form factors
are defined by:
 \begin{eqnarray}
  \langle A(P_2,\epsilon^*)|\bar q\gamma^{\mu}\gamma_5 b|\bar B(P_B)\rangle
   &=&-\frac{2iA(q^2)}{m_B-m_A}\epsilon^{\mu\nu\rho\sigma}
     \epsilon^*_{\nu}P_{B\rho}P_{2\sigma}, \nonumber\\
  \langle A(P_2,\epsilon^*)|\bar q\gamma^{\mu}b|\bar
  B(P_B)\rangle
   &=&-2m_A V_0(q^2)\frac{\epsilon^*\cdot q}{q^2}q^{\mu}
    -(m_B-m_A)V_1(q^2)\left[\epsilon^*_{\mu}
    -\frac{\epsilon^*\cdot q}{q^2}q^{\mu} \right] \nonumber\\
    &&+V_2(q^2)\frac{\epsilon^*\cdot q}{m_B-m_A}
     \left[ (P_B+P_2)^{\mu}-\frac{m_B^2-m_A^2}{q^2}q^{\mu} \right],\nonumber\\
  \langle A(P_2,\epsilon^*)|\bar q\sigma^{\mu\nu}\gamma_5q_{\nu}b|\bar
  B(P_B\rangle
   &=&-2T_1(q^2)\epsilon^{\mu\nu\rho\sigma}
     \epsilon^*_{\nu}P_{B\rho}P_{2\sigma}, \nonumber\\
  \langle A(P_2,\epsilon^*)|\bar q\sigma^{\mu\nu}q_{\nu}b|\bar
  B(P_B)\rangle
   &=&-iT_2(q^2)\left[(m_B^2-m_A^2)\epsilon^{*\mu}
       -(\epsilon^*\cdot q)(P_B+P_2)^{\mu} \right]\nonumber\\
   &&-iT_3(q^2)(\epsilon^*\cdot q)\left[
       q^{\mu}-\frac{q^2}{m_B^2-m_A^2}(P_B+P_2)^{\mu}\right],
 \end{eqnarray}
with a factor $-i$ different from $B\to V$ and the factor $m_B+m_V$
($m_B-m_V$) is replaced by $m_B-m_A$ ($m_B+m_A$). Similar to $B\to
V$ form factors, the relation $2m_AV_0=(m_B-m_A)V_1-(m_B+m_A)V_2$ is
obtained at $q^2=0$. In the PQCD approach, $B\to A$ form factors'
formulae can be derived from the corresponding $B\to V$ form factor
formulas using the replacement in Eq.~(\ref{eq:relationofAVDA}) with
the proper change of the sign of the pre-factor $r_2$ in $V$ and
$A_1$. The form factors in the large recoiling region can be
directly calculated.  In order to extrapolate the form factors to
the whole kinematic region, we use the results obtained in the
region $0< q^2<10\rm{GeV}$ and  we recast the form factors by
adopting the dipole parametrization for the form factors£º
 \begin{eqnarray}
 F(q^2)=\frac{F(0)}{1-a(q^2/m_B^2)+b(q^2/m_B^2)^2}\;.
 \end {eqnarray}

%%%%%%%%%%%%%%%%%%%%%%%%%%%%%%%%%%%%%%%%%%%%%%%%%%%%%%%%%%

\begin{table}
\caption{$B\to a_1,b_1,K_{1A,1B},h_1,h_8,f_1,f_8$ form factors.
$a,b$ are the parameters of the form factors in dipole
parametrization. The errors are from: decay constants of $B$ meson
and shape parameter $\omega_b$; $\Lambda_{\rm{QCD}}$ and the scales
$t_e$s; Gegenbauer moments of axial-vectors' LCDAs. }
 \label{Tab:formfactorsBtoAbeforemixing}
 \begin{center}
 \begin{tabular}{|c|c c c|c|c c c|}
\hline \hline
 $F$       & $F(0)$  & $a$ &$b$       & $F$       & $F(0)$  & $a$ &$b$                \\
 \hline
\hline
  $A^{B a_1}$    &$0.26_{-0.05-0.01-0.03}^{+0.06+0.00+0.03}$    &$1.72_{-0.05}^{+0.05}$    &$0.66_{-0.06}^{+0.07}$  &$A^{B b_1}$ & $0.19_{-0.04-0.01-0.03}^{+0.04+0.01+0.03}$      &$1.75_{-0.05}^{+0.06}$  &$0.70_{-0.05}^{+0.08}$  \\
 \hline
  $V_0^{B a_1}$  &$0.34_{-0.07-0.02-0.08}^{+0.07+0.01+0.08}$    &$1.73_{-0.06}^{+0.05}$    &$0.66_{-0.08}^{+0.06}$  &$V_0^{B b_1}$  & $0.45_{-0.09-0.01-0.04}^{+0.10+0.01+0.04}$    &$1.69_{-0.05}^{+0.05}$    &$0.61_{-0.07}^{+0.05}$  \\
 \hline
  $V_1^{B a_1}$  &$0.43_{-0.09-0.01-0.05}^{+0.10+0.01+0.05}$    &$0.75_{-0.05}^{+0.05}$    &$-0.12_{-0.02}^{+0.05}$  &$V_1^{B b_1}$  & $0.33_{-0.06-0.02-0.05}^{+0.07+0.01+0.05}$    &$0.80_{-0.06}^{+0.05}$    &$-0.09_{-0.05}^{+0.03}$  \\
 \hline
   $V_2^{B a_1}$   &$0.13_{-0.03-0.01-0.00}^{+0.03+0.00+0.00}$  &$--$  &$--$  &$V_2^{B b_1}$      &$0.03_{-0.01-0.00-0.02}^{+0.01+0.00+0.02}$  &$--$  &$--$\\
 \hline
 $T_1^{B a_1}$  &$0.34_{-0.07-0.01-0.05}^{+0.08+0.00+0.05}$    &$1.69_{-0.05}^{+0.06}$    &$0.61_{-0.05}^{+0.08}$  &$T_1^{B b_1}$   &$0.27_{-0.05-0.01-0.04}^{+0.06+0.01+0.04}$    &$1.70_{-0.06}^{+0.06}$    &$0.63_{-0.07}^{+0.07}$  \\
 \hline
 $T_2^{B a_1}$  &$0.34_{-0.07-0.01-0.05}^{+0.08+0.00+0.05}$    &$0.71_{-0.05}^{+0.07}$    &$-0.16_{-0.02}^{+0.03}$  &$T_2^{B b_1}$   &$0.27_{-0.05-0.01-0.04}^{+0.06+0.01+0.04}$    &$0.75_{-0.05}^{+0.06}$    &$-0.14_{-0.07}^{+0.08}$  \\
 \hline
  $T_3^{B a_1}$ &$0.30_{-0.06-0.01-0.05}^{+0.07+0.05+0.05}$    &$1.60_{-0.05}^{+0.06}$    &$0.53_{-0.04}^{+0.06}$  &$T_3^{B b_1}$   &$0.18_{-0.04-0.01-0.03}^{+0.04+0.01+0.03}$    &$1.41_{-0.07}^{+0.08}$    &$0.43_{-0.05}^{+0.07}$   \\
 \hline\hline
 $A^{B K_{1A}}$      &$0.27_{-0.05-0.01-0.06}^{+0.06+0.00+0.06}$    &$1.73_{-0.06}^{+0.07}$    &$0.67_{-0.07}^{+0.09}$  &$A^{B K_{1B}}$   &$0.20_{-0.04-0.01-0.05}^{+0.04+0.01+0.05}$    &$1.73_{-0.06}^{+0.07}$    &$0.68_{-0.06}^{+0.08}$   \\
\hline
 $V_0^{B K_{1A}}$    &$0.35_{-0.07-0.02-0.13}^{+0.08+0.01+0.13}$    &$1.73_{-0.09}^{+0.07}$    &$0.66_{-0.10}^{+0.09}$  &$V_0^{B K_{1B}}$   &$0.52_{-0.10-0.02-0.07}^{+0.12+0.01+0.07}$    &$1.72_{-0.06}^{+0.06}$    &$0.64_{-0.06}^{+0.07}$  \\
\hline
 $V_1^{B K_{1A}}$    &$0.47_{-0.09-0.01-0.01}^{+0.11+0.01+0.01}$    &$0.75_{-0.04}^{+0.09}$    &$-0.13_{-0.00}^{+0.10}$  &$V_1^{B K_{1B}}$   &$0.36_{-0.07-0.02-0.08}^{+0.08+0.01+0.09}$    &$0.78_{-0.05}^{+0.06}$    &$-0.10_{-0.03}^{+0.05}$  \\
\hline
  $V_2^{B K_{1A}}$   &$0.14_{-0.03-0.01-0.02}^{+0.03+0.00+0.02}$  &$--$  &$--$   &$V_2^{B K_{1B}}$    &$0.00_{-0.00-0.00-0.03}^{+0.00+0.00+0.03}$  &$--$  &$--$       \\
\hline
 $T_1^{B K_{1A}}$   &$0.37_{-0.07-0.01-0.01}^{+0.08+0.01+0.01}$    &$1.70_{-0.07}^{+0.08}$    &$0.63_{-0.09}^{+0.08}$  &$T_1^{B K_{1B}}$   &$0.29_{-0.06-0.01-0.06}^{+0.06+0.01+0.06}$     &$1.68_{-0.07}^{+0.08}$    &$0.61_{-0.06}^{+0.10}$  \\
\hline
 $T_2^{B K_{1A}}$   &$0.37_{-0.07-0.01-0.01}^{+0.08+0.01+0.01}$    &$0.72_{-0.07}^{+0.10}$    &$-0.16_{-0.01}^{+0.06}$  &$T_2^{B K_{1B}}$   &$0.29_{-0.06-0.01-0.06}^{+0.06+0.01+0.06}$      &$0.73_{-0.07}^{+0.07}$    &$-0.14_{-0.04}^{+0.03}$   \\
\hline
 $T_3^{B K_{1A}}$   &$0.33_{-0.07-0.01-0.08}^{+0.08+0.00+0.08}$    &$1.61_{-0.06}^{+0.09}$    &$0.54_{-0.05}^{+0.11}$  &$T_3^{B K_{1B}}$   &$0.20_{-0.04-0.01-0.05}^{+0.05+0.01+0.05}$      &$1.38_{-0.09}^{+0.08}$    &$.43_{-0.07}^{+0.06}$  \\
 \hline \hline
 $A^{B h_1}$   &$0.12_{-0.02-0.01-0.02}^{+0.03+0.00+0.02}$     &$1.73_{-0.05}^{+0.06}$    &$0.68_{-0.06}^{+0.08}$  &$A^{B h_8}$       &$0.09_{-0.02-0.00-0.01}^{+0.02+0.00+0.01}$     &$1.74_{-0.05}^{+0.06}$    &$0.68_{-0.05}^{+0.05}$  \\
\hline
 $V_0^{B h_1}$   &$0.26_{-0.05-0.01-0.02}^{+0.06+0.00+0.02}$    &$1.68_{-0.04}^{+0.06}$    &$0.59_{-0.04}^{+0.08}$  &$V_0^{B h_8}$    &$0.21_{-0.04-0.01-0.02}^{+0.05+0.00+0.02}$      &$1.70_{-0.06}^{+0.04}$    &$0.62_{-0.08}^{+0.04}$  \\
\hline
 $V_1^{B h_1}$   &$0.20_{-0.04-0.01-0.03}^{+0.04+0.01+0.03}$     &$0.77_{-0.04}^{+0.07}$    &$-0.11_{-0.01}^{+0.06}$  &$V_1^{B h_8}$   &$0.16_{-0.03-0.01-0.02}^{+0.04+0.01+0.02}$      &$0.78_{-0.05}^{+0.05}$    &$-0.10_{-0.06}^{+0.01}$  \\
\hline
 $V_2^{B h_1}$   &$0.03_{-0.00-0.00-0.01}^{+0.01+0.00+0.01}$  &$--$  &$--$  &$V_2^{B h_8}$    &$0.01_{-0.00-0.00-0.01}^{+0.00+0.00+0.01}$  &$--$  &$--$      \\
\hline
 $T_1^{B h_1}$   &$0.17_{-0.03-0.01-0.02}^{+0.04+0.00+0.02}$    &$1.69_{-0.06}^{+0.06}$    &$0.63_{-0.08}^{+0.06}$  &$T_1^{B h_8}$   &$0.13_{-0.03-0.01-0.02}^{+0.03+0.00+0.02}$      &$1.71_{-0.07}^{+0.05}$    &$0.65_{-0.11}^{+0.04}$   \\
\hline
 $T_2^{B h_1}$   &$0.17_{-0.03-0.01-0.02}^{+0.04+0.00+0.02}$    &$0.73_{-0.07}^{+0.05}$    &$-0.13_{-0.06}^{+0.01}$  &$T_2^{B h_8}$  &$0.13_{-0.03-0.01-0.02}^{+0.03+0.00+0.02}$      &$0.74_{-0.06}^{+0.05}$    &$-0.13_{-0.04}^{+0.00}$   \\
\hline
 $T_3^{B h_1}$   &$0.12_{-0.02-0.01-0.02}^{+0.03+0.01+0.02}$    &$1.41_{-0.08}^{+0.06}$    &$0.45_{-0.11}^{+0.01}$  &$T_3^{B h_8}$   &$0.09_{-0.02-0.01-0.01}^{+0.02+0.00+0.01}$      &$1.40_{-0.08}^{+0.07}$    &$0.44_{-0.07}^{+0.03}$   \\
 \hline \hline
 $A^{B f_1}$     &$0.16_{-0.03-0.00-0.02}^{+0.04+0.00+0.02}$    &$1.73_{-0.05}^{+0.05}$    &$0.67_{-0.07}^{+0.04}$  &$A^{B f_8}$    &$0.11_{-0.02-0.00-0.01}^{+0.03+0.00+0.01}$      &$1.72_{-0.05}^{+0.05}$    &$0.66_{-0.05}^{+0.07}$   \\
\hline
 $V_0^{B f_1}$   &$0.21_{-0.04-0.01-0.05}^{+0.05+0.01+0.05}$    &$1.73_{-0.05}^{+0.06}$    &$0.66_{-0.04}^{+0.08}$  &$V_0^{B f_8}$   &$0.15_{-0.03-0.01-0.03}^{+0.03+0.01+0.03}$     &$1.74_{-0.06}^{+0.05}$    &$0.68_{-0.07}^{+0.06}$   \\
\hline
 $V_1^{B f_1}$   &$0.27_{-0.05-0.01-0.03}^{+0.06+0.00+0.03}$    &$0.75_{-0.05}^{+0.05}$    &$-0.12_{-0.02}^{+0.05}$  &$V_1^{B f_8}$    &$0.19_{-0.04-0.00-0.02}^{+0.04+0.00+0.02}$      &$0.75_{-0.05}^{+0.05}$    &$-0.12_{-0.02}^{+0.04}$   \\
\hline
 $V_2^{B f_1}$   &$0.08_{-0.02-0.00-0.00}^{+0.02+0.00+0.00}$  &$--$  &$--$    &$V_2^{B f_8}$          &$0.05_{-0.01-0.00-0.00}^{+0.01+0.00+0.00}$  &$--$  &$--$     \\
\hline
 $T_1^{B f_1}$   &$0.21_{-0.04-0.00-0.03}^{+0.05+0.00+0.03}$    &$1.69_{-0.06}^{+0.06}$    &$0.62_{-0.06}^{+0.08}$  &$T_1^{B f_8}$   &$0.15_{-0.03-0.00-0.02}^{+0.03+0.00+0.02}$      &$1.68_{-0.05}^{+0.06}$    &$0.60_{-0.05}^{+0.07}$   \\
\hline
 $T_2^{B f_1}$   &$0.21_{-0.04-0.00-0.03}^{+0.05+0.00+0.03}$    &$0.72_{-0.07}^{+0.06}$    &$-0.15_{-0.06}^{+0.01}$  &$T_2^{B f_8}$   &$0.15_{-0.03-0.00-0.02}^{+0.03+0.00+0.02}$      &$0.71_{-0.05}^{+0.06}$    &$-0.16_{-0.03}^{+0.03}$   \\
\hline
 $T_3^{B f_1}$   &$0.19_{-0.04-0.00-0.03}^{+0.04+0.00+0.03}$    &$1.61_{-0.07}^{+0.05}$    &$0.55_{-0.09}^{+0.05}$  &$T_3^{B f_8}$   &$0.13_{-0.03-0.00-0.02}^{+0.03+0.00+0.02}$      &$1.61_{-0.05}^{+0.05}$    &$0.54_{-0.06}^{+0.06}$   \\
 \hline \hline
 \end{tabular}
 \end{center}
 \end{table}
%%%%%%%%%%%%%%%%%%%%%%%%%%%%%%%%%%%%%%%%%%%%%%%%%%%%%%%%%%

%%%%%%%%%%%%%%%%%%%%%%%%%%%%%%%%%%%%%%%%%%%%%%%%%%%%%%%%%%

\begin{table}
\caption{Same as Table \ref{Tab:formfactorsBtoAbeforemixing} except
$B_s\to h_8,h_1,f_8,f_1,K_{1A},K_{1B}$.}
 \label{Tab:formfactorsBstoAbeforemixing}
 \begin{center}
 \begin{tabular}{|c|c c c|c|c c c|}
\hline \hline
 $F$       & $F(0)$  & $a$ &$b$       & $F$       & $F(0)$  & $a$ &$b$                \\
 \hline \hline
 $A^{B_s K_{1A}}$     &$0.25_{-0.05-0.01-0.05}^{+0.05+0.00+0.05}$    &$1.73_{-0.07}^{+0.06}$    &$0.68_{-0.10}^{+0.04}$  &$A^{B_s K_{1B}}$      &$0.18_{-0.03-0.01-0.04}^{+0.04+0.00+0.04}$      &$.80_{-0.07}^{+0.06}$    &$0.76_{-0.10}^{+0.08}$   \\
\hline
 $V_0^{B_s K_{1A}}$   &$0.36_{-0.07-0.01-0.11}^{+0.07+0.00+0.11}$    &$1.76_{-0.07}^{+0.07}$    &$0.70_{-0.10}^{+0.09}$  &$V_0^{B_s K_{1B}}$     &$0.42_{-0.08-0.01-0.06}^{+0.09+0.01+0.06}$     &$1.69_{-0.05}^{+0.05}$    &$0.60_{-0.06}^{+0.06}$  \\
\hline
 $V_1^{B_s K_{1A}}$   &$0.43_{-0.08-0.01-0.09}^{+0.09+0.01+0.09}$    &$0.76_{-0.08}^{+0.06}$    &$-0.09_{-0.11}^{+0.00}$  &$V_1^{B_s K_{1B}}$    &$0.33_{-0.06-0.01-0.07}^{+0.07+0.00+0.07}$      &$0.84_{-0.05}^{+0.09}$    &$-0.09_{-0.01}^{+0.11}$   \\
\hline
 $V_2^{B_s K_{1A}}$       &$0.11_{-0.02-0.01-0.02}^{+0.02+0.01+0.02}$  &$--$  &$--$    &$V_2^{B_s K_{1B}}$        &$0.03_{-0.01-0.00-0.02}^{+0.01+0.00+0.02}$  &$--$  &$--$  \\
\hline
 $T_1^{B_s K_{1A}}$   &$0.34_{-0.06-0.01-0.07}^{+0.07+0.00+0.08}$    &$1.69_{-0.07}^{+0.07}$    &$0.62_{-0.10}^{+0.07}$  &$T_1^{B_s K_{1B}}$     &$0.26_{-0.05-0.01-0.06}^{+0.05+0.00+0.06}$      &$1.76_{-0.07}^{+0.06}$    &$0.71_{-0.09}^{+0.06}$    \\
\hline
 $T_2^{B_s K_{1A}}$   &$0.34_{-0.06-0.01-0.07}^{+0.07+0.00+0.08}$    &$0.71_{-0.07}^{+0.08}$    &$-0.15_{-0.05}^{+0.01}$  &$T_2^{B_s K_{1B}}$     &$0.26_{-0.05-0.01-0.06}^{+0.05+0.00+0.06}$      &$0.81_{-0.06}^{+0.08}$    &$-0.12_{-0.02}^{+0.08}$   \\
\hline
 $T_3^{B_s K_{1A}}$   &$0.30_{-0.06-0.01-0.07}^{+0.06+0.00+0.07}$    &$1.60_{-0.06}^{+0.06}$    &$0.54_{-0.07}^{+0.05}$  &$T_3^{B_s K_{1B}}$     &$0.17_{-0.03-0.01-0.04}^{+0.04+0.00+0.04}$      &$1.43_{-0.07}^{+0.10}$    &$0.44_{-0.02}^{+0.13}$    \\
 \hline \hline
 $A^{B_s h_1}$        &$0.10_{-0.02-0.00-0.02}^{+0.02+0.00+0.02}$    &$1.74_{-0.05}^{+0.06}$    &$0.69_{-0.06}^{+0.07}$  &$A^{B_s h_8}$           &$-0.16_{-0.03-0.00-0.02}^{+0.03+0.00+0.02}$      &$1.75_{-0.05}^{+0.06}$    &$0.70_{-0.05}^{+0.07}$   \\
\hline
 $V_0^{B_s h_1}$      &$0.23_{-0.04-0.00-0.02}^{+0.05+0.00+0.02}$    &$1.69_{-0.05}^{+0.05}$    &$0.61_{-0.06}^{+0.05}$  &$V_0^{B_s h_8}$         &$-0.36_{-0.07-0.00-0.03}^{+0.07+0.01+0.03}$      &$1.71_{-0.05}^{+0.04}$    &$0.63_{-0.05}^{+0.03}$   \\
\hline
 $V_1^{B_s h_1}$      &$0.18_{-0.03-0.00-0.03}^{+0.04+0.00+0.03}$    &$0.79_{-0.07}^{+0.05}$    &$-0.07_{-0.10}^{+0.01}$  &$V_1^{B_s h_8}$        &$-0.28_{-0.06-0.00-0.04}^{+0.05+0.01+0.04}$     &$0.79_{-0.05}^{+0.05}$    &$-0.08_{-0.05}^{+0.03}$    \\
\hline
 $V_2^{B_s h_1}$         &$0.03_{-0.00-0.00-0.01}^{+0.00+0.00+0.01}$  &$--$  &$--$      &$V_2^{B_s h_8}$         &$-0.02_{-0.00-0.00-0.01}^{+0.00+0.00+0.01}$  &$--$  &$--$  \\
\hline
 $T_1^{B_s h_1}$      &$0.15_{-0.03-0.00-0.02}^{+0.03+0.00+0.02}$    &$1.69_{-0.05}^{+0.06}$    &$0.63_{-0.06}^{+0.07}$  &$T_1^{B_s h_8}$      &$-0.23_{-0.05-0.00-0.03}^{+0.04+0.00+0.03}$      &$1.71_{-0.06}^{+0.05}$    &$0.65_{-0.09}^{+0.03}$   \\
\hline
 $T_2^{B_s h_1}$      &$0.15_{-0.03-0.00-0.02}^{+0.03+0.00+0.02}$    &$0.73_{-0.05}^{+0.07}$    &$-0.14_{-0.02}^{+0.05}$  &$T_2^{B_s h_8}$     &$-0.23_{-0.05-0.00-0.03}^{+0.04+0.00+0.03}$      &$0.75_{-0.06}^{+0.05}$    &$-0.13_{-0.03}^{+0.03}$    \\
\hline
 $T_3^{B_s h_1}$      &$0.10_{-0.02-0.00-0.01}^{+0.02+0.00+0.01}$     &$1.39_{-0.06}^{+0.07}$    &$0.42_{-0.03}^{+0.09}$  &$T_3^{B_s h_8}$      &$-0.15_{-0.03-0.00-0.02}^{+0.03+0.01+0.02}$      &$1.40_{-0.07}^{+0.07}$    &$0.43_{-0.04}^{+0.07}$    \\
 \hline \hline
 $A^{B_s f_1}$        &$0.14_{-0.03-0.00-0.02}^{+0.03+0.00+0.02}$    &$1.73_{-0.04}^{+0.06}$    &$0.66_{-0.04}^{+0.07}$  &$A^{B_s f_8}$        &$-0.19_{-0.04-0.00-0.02}^{+0.04+0.01+0.02}$      &$1.72_{-0.03}^{+0.06}$    &$0.65_{-0.02}^{+0.09}$    \\
\hline
 $V_0^{B_s f_1}$      &$0.18_{-0.03-0.00-0.04}^{+0.04+0.00+0.04}$    &$1.74_{-0.05}^{+0.06}$    &$0.68_{-0.06}^{+0.08}$  &$V_0^{B_s f_8}$      &$-0.26_{-0.05-0.00-0.05}^{+0.05+0.01+0.05}$      &$1.75_{-0.05}^{+0.06}$    &$0.69_{-0.05}^{+0.09}$    \\
\hline
 $V_1^{B_s f_1}$      &$0.23_{-0.04-0.01-0.03}^{+0.05+0.00+0.03}$    &$0.76_{-0.05}^{+0.05}$    &$-0.11_{-0.04}^{+0.03}$  &$V_1^{B_s f_8}$     &$-0.33_{-0.07-0.00-0.04}^{+0.06+0.01+0.04}$      &$0.76_{-0.07}^{+0.04}$    &$-0.09_{-0.13}^{+0.00}$   \\
\hline
 $V_2^{B_s f_1}$         &$0.07_{-0.01-0.00-0.00}^{+0.01+0.00+0.00}$  &$--$  &$--$     &$V_2^{B_s f_8}$         &$-0.10_{-0.02-0.00-0.00}^{+0.02+0.01+0.00}$  &$--$  &$--$ \\
\hline
 $T_1^{B_s f_1}$      &$0.18_{-0.03-0.01-0.03}^{+0.04+0.00+0.03}$    &$1.69_{-0.05}^{+0.06}$    &$0.62_{-0.06}^{+0.07}$  &$T_1^{B_s f_8}$     &$-0.26_{-0.05-0.00-0.03}^{+0.05+0.01+0.03}$      &$1.69_{-0.05}^{+0.06}$    &$0.61_{-0.06}^{+0.07}$     \\
\hline
 $T_2^{B_s f_1}$      &$0.18_{-0.03-0.01-0.03}^{+0.04+0.00+0.03}$    &$0.72_{-0.05}^{+0.07}$    &$-0.16_{-0.02}^{+0.04}$  &$T_2^{B_s f_8}$    &$-0.26_{-0.05-0.10-0.03}^{+0.05+0.01+0.03}$      &$0.71_{-0.05}^{+0.06}$    &$-0.15_{-0.04}^{+0.02}$   \\
\hline
 $T_3^{B_s f_1}$      &$0.16_{-0.03-0.00-0.03}^{+0.03+0.00+0.03}$    &$1.60_{-0.05}^{+0.06}$    &$0.54_{-0.05}^{+0.07}$  &$T_3^{B_s f_8}$     &$-0.23_{-0.05-0.00-0.03}^{+0.04+0.01+0.03}$      &$1.61_{-0.05}^{+0.05}$    &$0.55_{-0.07}^{+0.05}$    \\
 \hline \hline
 \end{tabular}
 \end{center}
 \end{table}
%%%%%%%%%%%%%%%%%%%%%%%%%%%%%%%%%%%%%%%%%%%%%%%%%%%%%%%%%%

%%%%%%%%%%%%%%%%%%%%%%%%%%%%%%%%%%%%%%%%%%%%%%%%%%%%%%%%%%
\begin{table}
\caption{$B\to a_1,b_1$ form factors at maximally recoil and the
results in the light-front quark model(LFQM) and light-cone sum
rules(LCSR). The errors in this work are from: decay constants of
$B$ meson and shape parameter $\omega_b$; $\Lambda_{\rm{QCD}}$ and
the scales $t_e$s; Gegenbauer moments in axial-vectors' LCDAs. }
 \label{Tab:threeCompare}
 \begin{center}
 \begin{tabular}{|cc|c|c|c|c|c|c}
\hline \hline
         &$B\to a_1$       & This work   & LFQM\cite{Cheng:2003sm,Cheng:2004yj}&         LCSR\cite{Yang:BtoAinLCSR}           \\
 \hline
\hline
\ \ \    & $A$        &$0.26_{-0.05-0.01-0.03}^{+0.06+0.00+0.03}$        &$0.25$       &$0.48\pm 0.09$            \\
\hline   & $V_0$     &$0.34_{-0.07-0.02-0.08}^{+0.07+0.01+0.08}$         &$0.13$       &$0.30\pm 0.05$           \\
\hline   & $V_1$     &$0.43_{-0.09-0.01-0.05}^{+0.10+0.01+0.05}$         &$0.37$       &$0.37\pm 0.07$           \\
\hline   & $V_2$     &$0.13_{-0.03-0.01-0.00}^{+0.03+0.00+0.00}$                                             &$0.18$       &$0.42\pm 0.08$           \\
\hline  &$T_1(T_2)$  &$0.34_{-0.07-0.01-0.05}^{+0.08+0.00+0.05}$         &$--$         &$--$            \\
\hline  & $T_3$      &$0.30_{-0.06-0.01-0.05}^{+0.07+0.05+0.05}$          &$--$        &$--$             \\
 \hline \hline
        & $B\to b_1$      & This work    & LFQM\cite{Cheng:2003sm,Cheng:2004yj}&         LCSR\cite{Yang:BtoAinLCSR}      \\
\hline
\ \ \   & $A$        &$0.19_{-0.04-0.01-0.03}^{+0.04+0.01+0.03}$    &$0.10$   &$-0.25\pm 0.05$   \\
\hline
\ \ \   & $V_0$    & $0.45_{-0.09-0.01-0.04}^{+0.10+0.01+0.04}$    &$0.39$    &$-0.39\pm 0.07$\\
\hline
\ \ \   & $V_1$    &$0.33_{-0.06-0.02-0.05}^{+0.07+0.01+0.05}$     &$0.18$    &$-0.20\pm 0.04$ \\
\hline
\ \ \   & $V_2$    &$0.03_{-0.01-0.00-0.02}^{+0.01+0.00+0.02}$                                         &$-0.03$   &$-0.09\pm 0.02$ \\%
\hline
\ \ \  &$T_1(T_2)$ &$0.27_{-0.05-0.01-0.04}^{+0.06+0.01+0.04}$     &$--$      &$--$\\
\hline
\ \ \  & $T_3$     &$0.18_{-0.04-0.01-0.03}^{+0.04+0.01+0.03}$     &$--$      &$--$\\
\hline      \hline
\end{tabular}
\end{center}
\end{table}
%%%%%%%%%%%%%%%%%%%%%%%%%%%%%%%%%%%%%%%%%%%%%%%%%%%%%%%%%%

%%%%%%%%%%%%%%%%%%%%%%%%%%%%%%%%%%%%%%%%%%%%%%%%%%%%%%%%%%

%%%%%%%%%%%%%%%%%%%%%%%%%%%%%%%%%%%%%%%%%%%%%%%%%%%%%%%%%%
{\small
\begin{table}
\caption{$B_{u,d,s}\to K_1(1270), K_1(1400)$, $B_{u,d,s}\to
h_1(1170), h_1(1380)$ and $B_{u,d,s}\to f_1(1285), h_1(1420)$ form
factors for physical axial-vector mesons at maximally recoil point,
i.e. $q^2=0$. Results in the first line of each form factor are
calculated using $\theta_K=45^\circ$, $\theta_{^1P_1}=10^\circ$ or
$\theta_{^3P_1}=38^\circ$, while the second line corresponds to the
angle $\theta_K=-45^\circ$, $\theta_{^1P_1}=45^\circ$ or
$\theta_{^3P_1}=50^\circ$. The errors are from: decay constants of
$B_{(s)}$ meson and shape parameter $\omega_b$; $\Lambda_{\rm{QCD}}$
and the scales $t_e$s; Gegenbauer moments in axial-vectors' LCDAs. }
 \label{Tab:formfactorsBtoAaftermixing}
\begin{center}
\begin{tabular}{|cc|c|c|c|c|}
\hline \hline
         &   $$        &$B\to K_1(1270)$    & $B\to K_1(1400)$&$B_s\to K_1(1270)$  & $B_s\to K_1(1400)$           \\
 \hline\hline
 \ \ \   & $A$         &$0.33_{-0.07-0.01-0.03}^{+0.08+0.07+0.03}$         &$0.05_{-0.01-0.00-0.07}^{+0.01+0.00+0.07}$      &$0.31_{-0.06-0.00-0.06}^{+0.06+0.00+0.06}$        &$0.05_{-0.01-0.01-0.02}^{+0.01+0.00+0.02}$       \\
 \ \ \   &$$           &$-0.05_{-0.01-0.00-0.07}^{+0.01+0.00+0.07}$        &$0.33_{-0.08-0.01-0.03}^{+0.07+0.01+0.03}$     &$-0.05_{-0.01-0.00-0.02}^{+0.01+0.01+0.02}$        &$0.31_{-0.06-0.00-0.06}^{+0.06+0.00+0.06}$       \\
 \hline   & $V_0$      &$0.62_{-0.12-0.03-0.06}^{+0.14+0.02+0.06}$         &$-0.12_{-0.02-0.00-0.14}^{+0.03+0.00+0.14}$     &$0.55_{-0.10-0.01-0.12}^{+0.11+0.01+0.12}$        &$-0.04_{-0.01-0.01-0.05}^{+0.01+0.01+0.05}$       \\
  \ \ \   &$$          &$0.12_{-0.02-0.00-0.14}^{+0.03+0.00+0.14}$         &$0.62_{-0.14-0.02-0.06}^{+0.12+0.03+0.06}$     &$0.04_{-0.01-0.01-0.05}^{+0.01+0.01+0.05}$         &$0.55_{-0.10-0.01-0.12}^{+0.11+0.01+0.12}$       \\
 \hline   & $V_1$      &$0.59_{-0.11-0.02-0.05}^{+0.13+0.01+0.05}$         &$0.08_{-0.02-0.01-0.12}^{+0.02+0.01+0.12}$      &$0.54_{-0.10-0.01-0.11}^{+0.11+0.01+0.11}$        &$0.07_{-0.01-0.01-0.04}^{+0.01+0.01+0.04}$        \\
   \ \ \   &$$         &$-0.08_{-0.02-0.01-0.12}^{+0.02+0.01+0.12}$        &$0.59_{-0.13-0.01-0.05}^{+0.11+0.02+0.05}$     &$-0.07_{-0.01-0.01-0.04}^{+0.01+0.01+0.04}$        &$0.54_{-0.11-0.01-0.11}^{+0.10+0.01+0.11}$       \\
 \hline   & $V_2$     &$0.11_{-0.02-0.00-0.01}^{+0.03+0.00+0.01}$         &$0.09_{-0.02-0.01-0.03}^{+0.02+0.00+0.03}$      &$0.12_{-0.02-0.00-0.03}^{+0.02+0.00+0.03}$        &$0.05_{-0.01-0.01-0.01}^{+0.01+0.01+0.01}$        \\
  \ \ \   &           &$-0.09_{-0.02-0.00-0.04}^{+0.02+0.01+0.04}$         &$0.07_{-0.01-0.00-0.01}^{+0.02+0.00+0.01}$      &$-0.06_{-0.01-0.01-0.01}^{+0.01+0.01+0.01}$        &$0.08_{-0.01-0.00-0.03}^{+0.02+0.00+0.03}$       \\
 \hline  &$T_1(T_2)$  &$0.46_{-0.09-0.02-0.04}^{+0.10+0.01+0.04}$          &$0.05_{-0.01-0.00-0.10}^{+0.01+0.01+0.10}$      &$0.43_{-0.08-0.01-0.09}^{+0.09+0.00+0.09}$        &$0.05_{-0.01-0.01-0.03}^{+0.01+0.00+0.03}$       \\
  \ \ \  &$$          &$-0.05_{-0.01-0.00-0.10}^{+0.01+0.01+0.10}$         &$0.46_{-0.10-0.01-0.04}^{+0.09+0.02+0.04}$     &$-0.05_{-0.01-0.00-0.03}^{+0.01+0.01+0.03}$        &$0.43_{-0.09-0.01-0.09}^{+0.08+0.01+0.09}$       \\
 \hline  & $T_3$      &$0.37_{-0.07-0.02-0.04}^{+0.09+0.01+0.04}$          &$0.09_{-0.02-0.00-0.08}^{+0.02+0.00+0.08}$      &$0.33_{-0.06-0.01-0.07}^{+0.07+0.00+0.07}$        &$0.10_{-0.02-0.01-0.03}^{+0.02+0.00+0.03}$\\
  \ \ \   &$$         &$-0.09_{-0.02-0.00-0.08}^{+0.02+0.00+0.08}$         &$0.37_{-0.09-0.01-0.04}^{+0.07+0.02+0.04}$     &$-0.10_{-0.02-0.00-0.03}^{+0.02+0.01+0.03}$        &$0.33_{-0.06-0.01-0.07}^{+0.07+0.00+0.07}$       \\
 \hline \hline
         &         &$B\to h_1(1170)$    &$B\to h_1(1380)$   & $B_s\to h_1(1170)$   & $B_s\to h_1(1380)$          \\
 \hline
\hline
\ \ \   & $A$            &$0.13_{-0.03-0.01-0.02}^{+0.03+0.00+0.02}$       &$0.07_{-0.01-0.00-0.01}^{+0.01+0.00+0.01}$          &$0.08_{-0.01-0.00-0.01}^{+0.02+0.00+0.01}$    &$-0.17_{-0.04-0.00-0.03}^{+0.03+0.00+0.03}$     \\
\ \ \   &                &$0.15_{-0.03-0.01-0.02}^{+0.03+0.01+0.02}$       &$-0.02_{-0.00-0.00-0.00}^{+0.00+0.00+0.00}$         &$-0.04_{-0.01-0.00-0.01}^{+0.01+0.00+0.01}$   &$-0.18_{-0.04-0.00-0.03}^{+0.03+0.01+0.03}$\\
\hline   & $V_0$         &$0.30_{-0.06-0.01-0.02}^{+0.07+0.01+0.02}$       &$0.16_{-0.03-0.01-0.01}^{+0.03+0.00+0.01}$          &$0.17_{-0.03-0.00-0.01}^{+0.03+0.00+0.01}$    &$-0.40_{-0.08-0.01-0.03}^{+0.07+0.01+0.03}$     \\
\ \ \   &                &$0.33_{-0.06-0.01-0.03}^{+0.07+0.01+0.03}$       &$-0.04_{-0.01-0.00-0.00}^{+0.01+0.00+0.00}$         &$-0.09_{-0.02-0.00-0.01}^{+0.02+0.00+0.01}$   &$-0.42_{-0.09-0.01-0.03}^{+0.08+0.01+0.03}$\\
\hline   & $V_1$         &$0.23_{-0.04-0.01-0.03}^{+0.05+0.01+0.03}$       &$0.12_{-0.02-0.01-0.02}^{+0.03+0.00+0.02}$          &$0.13_{-0.02-0.00-0.02}^{+0.03+0.00+0.02}$    &$-0.31_{-0.06-0.00-0.05}^{+0.06+0.01+0.05}$        \\
\ \ \   &                &$0.26_{-0.05-0.01-0.04}^{+0.06+0.01+0.04}$       &$-0.03_{-0.01-0.00-0.00}^{+0.01+0.00+0.00}$         &$-0.07_{-0.02-0.00-0.01}^{+0.01+0.00+0.01}$   &$-0.32_{-0.07-0.00-0.05}^{+0.06+0.01+0.05}$\\
\hline   & $V_2$        &$0.04_{-0.01-0.00-0.01}^{+0.01+0.00+0.01}$         &$0.00_{-0.00-0.00-0.01}^{+0.00+0.00+0.01}$      &$0.02_{-0.00-0.00-0.01}^{+0.00+0.00+0.01}$       &$-0.02_{-0.00-0.00-0.02}^{+0.00+0.00+0.02}$        \\
\ \ \   &               &$0.04_{-0.01-0.00-0.02}^{+0.01+0.00+0.02}$         &$-0.00_{-0.00-0.00-0.00}^{+0.00+0.00+0.00}$      &$-0.01_{-0.00-0.00-0.01}^{+0.00+0.00+0.01}$     &$-0.02_{-0.00-0.00-0.02}^{+0.00+0.00+0.02}$        \\
\hline  &$T_1(T_2)$      &$0.19_{-0.04-0.01-0.03}^{+0.04+0.01+0.03}$       &$0.10_{-0.02-0.00-0.01}^{+0.02+0.00+0.01}$          &$0.10_{-0.02-0.00-0.01}^{+0.02+0.00+0.01}$    &$-0.25_{-0.05-0.00-0.04}^{+0.05+0.01+0.04}$     \\
\ \ \   &                &$0.20_{-0.04-0.01-0.03}^{+0.05+0.01+0.03}$       &$-0.03_{-0.01-0.00-0.00}^{+0.00+0.00+0.00}$         &$-0.06_{-0.01-0.00-0.01}^{+0.01+0.00+0.01}$   &$-0.26_{-0.05-0.00-0.04}^{+0.05+0.01+0.04}$\\
\hline  & $T_3$          &$0.13_{-0.03-0.01-0.02}^{+0.03+0.01+0.02}$       &$0.07_{-0.01-0.00-0.01}^{+0.02+0.00+0.01}$          &$0.07_{-0.01-0.00-0.01}^{+0.02+0.00+0.01}$    &$-0.17_{-0.04-0.00-0.02}^{+0.03+0.01+0.02}$  \\
\ \ \   &                &$0.14_{-0.03-0.01-0.02}^{+0.03+0.01+0.02}$       &$-0.02_{-0.00-0.00-0.00}^{+0.00+0.00+0.00}$         &$-0.04_{-0.01-0.00-0.01}^{+0.01+0.00+0.01}$   &$-0.18_{-0.04-0.00-0.03}^{+0.03+0.01+0.03}$\\
 \hline \hline
         &              &$B\to f_1(1285)$     &$B\to f_1(1420)$      &$B_s\to f_1(1285)$   &$B_s\to f_1(1420)$\\
 \hline
\ \ \   & $A$             &$0.19_{-0.04-0.00-0.02}^{+0.04+0.00+0.02}$      &$-0.01_{-0.00-0.00-0.00}^{+0.00+0.00+0.00}$    &$-0.01_{-0.011-0.00-0.00}^{+0.00+0.00+0.00}$   &$-0.24_{-0.05-0.00-0.03}^{+0.04+0.00+0.03}$ \\
\ \ \   &                 &$0.18_{-0.04-0.00-0.02}^{+0.04+0.00+0.02}$      &$-0.05_{-0.01-0.00-0.01}^{+0.01+0.00+0.01}$    &$-0.06_{-0.01-0.00-0.00}^{+0.01+0.00+0.00}$   &$-0.23_{-0.05-0.00-0.03}^{+0.04+0.00+0.03}$\\
\hline  & $V_0$           &$0.26_{-0.05-0.01-0.06}^{+0.06+0.01+0.06}$      &$-0.01_{-0.00-0.00-0.01}^{+0.00+0.00+0.01}$    &$-0.02_{-0.00-0.00-0.00}^{+0.00+0.00+0.00}$   &$-0.32_{-0.06-0.00-0.07}^{+0.06+0.01+0.07}$\\
\ \ \   &                 &$0.25_{-0.05-0.01-0.06}^{+0.05+0.01+0.06}$      &$-0.07_{-0.01-0.00-0.02}^{+0.01+0.00+0.02}$    &$-0.08_{-0.02-0.00-0.00}^{+0.02+0.00+0.00}$   &$-0.31_{-0.06-0.04-0.07}^{+0.06+0.01+0.07}$\\
\hline   & $V_1$          &$0.33_{-0.07-0.01-0.04}^{+0.07+0.00+0.04}$      &$-0.02_{-0.00-0.0.-0.00}^{+0.00+0.00+0.00}$    &$-0.02_{-0.00-0.00-0.00}^{+0.00+0.00+0.00}$   &$-0.40_{-0.08-0.01-0.05}^{+0.07+0.01+0.05}$\\
\ \ \   &                 &$0.32_{-0.06-0.01-0.04}^{+0.07+0.00+0.04}$      &$-0.08_{-0.02-0.00-0.01}^{+0.02+0.00+0.01}$    &$-0.10_{-0.02-0.00-0.01}^{+0.02+0.00+0.01}$   &$-0.39_{-0.08-0.01-0.05}^{+0.07+0.01+0.05}$\\
\hline   & $V_2$         &$0.09_{-0.01-0.00-0.00}^{+0.02+0.00+0.00}$         &$-0.01_{-0.00-0.00-0.00}^{+0.00+0.00+0.00}$      &$-0.00_{-0.00-0.00-0.00}^{+0.00+0.00+0.00}$        &$-0.09_{-0.02-0.00-0.00}^{+0.02+0.01+0.00}$           \\
\ \ \   &                &$0.09_{-0.01-0.00-0.00}^{+0.02+0.00+0.00}$         &$-0.02_{-0.00-0.00-0.00}^{+0.00+0.00+0.00}$      &$-0.03_{-0.01-0.00-0.00}^{+0.01+0.00+0.00}$        &$-0.09_{-0.02-0.00-0.00}^{+0.02+0.01+0.00}$           \\
\hline  &$T_1(T_2)$       &$0.26_{-0.05-0.01-0.03}^{+0.06+0.00+0.03}$      &$-0.01_{-0.00-0.00-0.00}^{+0.00+0.00+0.00}$    &$-0.02_{-0.00-0.00-0.00}^{+0.00+0.00+0.00}$   &$-0.31_{-0.07-0.00-0.04}^{+0.06+0.00+0.04}$\\
\ \ \   &                 &$0.25_{-0.05-0.01-0.03}^{+0.06+0.00+0.03}$      &$-0.07_{-0.01-0.00-0.01}^{+0.01+0.00+0.01}$    &$-0.08_{-0.02-0.00-0.00}^{+0.02+0.00+0.00}$   &$-0.30_{-0.06-0.00-0.04}^{+0.06+0.01+0.04}$\\
\hline   &$T_3$           &$0.23_{-0.05-0.01-0.03}^{+0.05+0.00+0.03}$      &$-0.01_{-0.00-0.00-0.00}^{+0.00+0.00+0.00}$    &$-0.01_{-0.00-0.00-0.00}^{+0.00+0.00+0.00}$   &$-0.28_{-0.06-0.00-0.04}^{+0.05+0.01+0.04}$\\
\ \ \   &                 &$0.22_{-0.04-0.01-0.03}^{+0.05+0.00+0.03}$      &$-0.06_{-0.01-0.00-0.01}^{+0.01+0.00+0.01}$    &$-0.07_{-0.02-0.00-0.01}^{+0.02+0.00+0.01}$   &$-0.27_{-0.06-0.00-0.04}^{+0.05+0.01+0.04}$\\
\hline\hline
\end{tabular}
\end{center}
\end{table}}
%%%%%%%%%%%%%%%%%%%%%%%%%%%%%%%%%%%%%%%%%%%%%%%%%%%%%%%%%%

The real physical states $K_1(1270)$ and $K_1(1400)$ are mixtures of
the $K_{1A}$ and $K_{1B}$ states with the mixing angle $\theta_K$:
\begin{eqnarray}
|K_1(1270)\rangle&=&|K_{1A}\rangle
{\rm{sin}}\theta_K+|K_{1B}\rangle{\rm{cos}}\theta_K,\\
|K_1(1400)\rangle&=&|K_{1A}\rangle
{\rm{cos}}\theta_K-|K_{1B}\rangle{\rm{sin}}\theta_K.
\end{eqnarray}
In the flavor SU(3) symmetry limit, these mesons can not mix with
each other; but since  $s$ quark is heavier than the $u,d$ quarks,
$K_1(1270)$ and $K_1(1400)$ are not purely $1^3P_1$ or $1^1P_1$
states. Generally, the mixing angle can be determined by the
experimental data. One ideal method is making use of the decay
$\tau^-\to K_1\nu_\tau$, whose partial decay rate is given by
\begin{eqnarray}
{\Gamma}(\tau^-\to
K_1\nu_\tau)=\frac{m_\tau^3}{16\pi}G_F^2|V_{us}|^2f_A^2\left(1-\frac{m_A^2}{m_\tau^2}
\right)^2\left(1+\frac{2m_A^2}{m_\tau^2}\right),
\end{eqnarray}
with the measured results for branching
fractions~\cite{Amsler:2008zz}:
\begin{eqnarray}
{\cal BR}(\tau^-\to K_1(1270)\nu_\tau)=(4.7\pm1.1)\times 10^{-3},\;
{\cal BR}(\tau^-\to K_1(1400)\nu_\tau)=(1.7\pm2.6)\times
10^{-3}.\label{eq:BRintau}
\end{eqnarray}
The longitudinal decay constants (in MeV) can be straightly
obtained:
\begin{eqnarray}
|f_{K_1(1270)}|=169_{-21}^{+19};\;\;\;
|f_{K_1(1400)}|=125_{-125}^{+~74}.\label{eq:decayconstantK1}
\end{eqnarray}
In principle, one can combine the decay constants for $K_{1A}$,
$K_{1B}$ evaluated in QCD sum rules with the above results to
determine the mixing angle $\theta_K$. But since there are large
uncertainties in Eq.~(\ref{eq:decayconstantK1}), the constraint on
the mixing angle is expected to be rather smooth:
\begin{eqnarray}
 -143^\circ<\theta_K<-120^\circ,\;\;\;{\rm or}\;\;
 -49^\circ<\theta_K<-27^\circ,\;\;\;{\rm or}\;\;
 37^\circ<\theta_K<60^\circ,\;\;\;{\rm or}\;\;
 131^\circ<\theta_K<153^\circ,
\end{eqnarray}
where we have taken the uncertainties from the branching ratios in
Eq.(\ref{eq:BRintau}) and the first Gegenbauer moment $a_1^{K_1}$
into account but neglected the mass differences as usual. In this
paper, for simplicity, we use two reference values in
Ref.~\cite{Yang:2007zt}
\begin{eqnarray}
\theta_K=\pm 45^\circ.
\end{eqnarray}
Besides, the flavor-octet and the flavor-singlet also mix with each
other:
\begin{eqnarray}
 |f_1(1285)\rangle&=&|f_{1}\rangle
{\rm{cos}}\theta_{^3P_1}+|f_{8}\rangle{\rm{sin}}\theta_{^3P_1},\;\;\;
 |f_1(1420)\rangle=-|f_{1}\rangle
{\rm{sin}}\theta_{^3P_1}+|f_{8}\rangle{\rm{cos}}\theta_{^3P_1},\\
 |h_1(1170)\rangle&=&|h_{1}\rangle
{\rm{cos}}\theta_{^1P_1}+|h_{8}\rangle{\rm{sin}}\theta_{^1P_1},\;\;\;
 |h_1(1380)\rangle=-|h_{1}\rangle
{\rm{sin}}\theta_{^1P_1}+|h_{8}\rangle{\rm{cos}}\theta_{^1P_1}.
\end{eqnarray}
The reference points are chosen as: $\theta_{^3P_1}=38^\circ$ or
$\theta_{^3P_1}=50^\circ$; $\theta_{^1P_1}=10^\circ$ or
$\theta_{^1P_1}=45^\circ$ \cite{Yang:2007zt}. These reference points
are very close to the ideal mixing angle
$\theta_{^3P_1}=35.3^\circ$.  We should point out that if the mixing
is ideal: $f_1(1285)$ is made up of $\frac{\bar uu+\bar dd}{\sqrt
2}$ while $f_1(1420)$ is composed of $\bar ss$. As a result, some of
the form factors are very small, which leads to small production
rates of this meson.

In Table~\ref{Tab:formfactorsBtoAbeforemixing} and
\ref{Tab:formfactorsBstoAbeforemixing}, results of the form factors
at $q^2=0$ for $B_{u,d,s}\to
a_1,f_1,f_8,K_{1A},b_1,h_1,h_8\;{\rm{and}}\;K_{1B}$ transitions are
listed, together with the parameters $a,b$, which are obtained
within the dipole parametrization. The form factors for the
$B_{(s)}$ to physical states transitions are collected in Table
\ref{Tab:formfactorsBtoAaftermixing}. In our calculation, minus
values for decays constants of $^1P_1$ mesons~\footnote{Decay
constants given in QCD sum rules \cite{Yang:2005gk,Yang:2007zt} are
both positive for two kinds of axial-vectors, which will give
negative values for $B\to ^1P_1$ form factors. For non-strange
$^1P_1$ mesons, this minus sign will not give any physical
differences as it can not be observed experimentally. But we should
point out that the minus sign will affect the mixing between
$K_{1A}$ and $K_{1B}$ by changing the mixing angle $\theta$ to
$-\theta$.} have been used. The errors in the results are from:
decay constants of $B_{(s)}$ mesons and shape parameters $\omega_b$;
$\Lambda_{\rm{QCD}}\big((0.25\pm0.05)\rm{GeV}\big)$ and the scales
$t_e$s; Gegenbauer moments of axial-vectors' LCDAs. As the quark
contents (to be more precise, the mixing angles) of the
axial-vectors $K_1(f_1,h_1)$ have not been uniquely determined, we
give two sets of results for form factors as in
Ref.~\cite{Wang:2007an}: in Table
\ref{Tab:formfactorsBtoAaftermixing}, the results in the first line
are obtained using $\theta_K=45^\circ$, $\theta_{^1P_1}=10^\circ$
and $\theta_{^3P_1}=38^\circ$ while the second line using
$\theta_K=-45^\circ$, $\theta_{^1P_1}=45^\circ$ and
$\theta_{^3P_1}=50^\circ$.

%%%%%%%%%%%%%%%%%%%%%%%%%%%%%%%%%%%%%%%%%%%%%%%%%%%%%%%%%%
\begin{table}[tb]
\caption{Contributions to form factor $A_0$ and $T_1$ from different
distribution amplitudes.}
 \label{Tab:formfactorcomparison}
\begin{center}
 \begin{tabular}{c|c|c|c}
  \hline\hline
        $A_0$    &  $B\to\rho$ & $B\to a_1(1260)$    & $B\to b_1(1235)$
          \\  \hline
   $\phi$
                                                          &$0.108$
                                                          &$0.102$
                                                          &$0.199$
                                                          \\\hline
   $\phi^s$
                                                          &$0.103$
                                                          &$0.155$
                                                          &$0.179$
                                                          \\ \hline
   $\phi^t$
                                                          &$0.040$
                                                          &$0.081$
                                                          &$0.069$
                                                           \\
 \hline
 total                                                    &$0.251$
                                                          &$0.338$
                                                          &$0.446$
                                                          \\%
 \hline\hline
        $T_1$    &  $B\to\rho$ & $B\to a_1(1260)$    & $B\to b_1(1235)$
          \\  \hline
   $\phi^T$
                                                          &$0.085$
                                                          &$0.140$
                                                          &$0.082$
                                                          \\\hline
   $\phi^a$
                                                          &$0.047$
                                                          &$0.084$
                                                          &$0.085$
                                                          \\ \hline
   $\phi^v$
                                                          &$0.063$
                                                          &$0.113$
                                                          &$0.099$
                                                           \\
 \hline
 total                                                    &$0.194$
                                                          &$0.337$
                                                          &$0.266$
                                                          \\
 \hline\hline
\end{tabular}
\end{center}
\end{table}
%%%%%%%%%%%%%%%%%%%%%%%%%%%%%%%%%%%%%%%%%%%%%%%%%%%%%%%%%%

A number of remarks on $B\to A$ form factors are given in order.
\begin{enumerate}

\item The parameters $a$ in most form factors are around $1.7$,
but these parameters in $V_1(q^2)$ and $T_2(q^2)$ are around $0.7$.
The situation is similar for the parameter $b$. In most form
factors, this parameter is close to $0.7$, while in $V_1(q^2)$ and
$T_2(q^2)$ it's close to $-0.14$.

\item Some of the form factors for the two kinds of axial-vector mesons are very different.
As an example,  we will give a comparison of the $B\to\rho$, $B\to
a_1(1260)$ and $B\to b_1(1235)$ form factors. Form factors $V_0$,
$V_1$, $T_1$ for $B\to A$ transition are larger than the
corresponding $B\to V$ ones. It seems that the form factor $A^{B\to
(a_1,b_1)}$ is somewhat equal to or even smaller than $V^{B\to
\rho}$. But actually that is artificial: as in
Eq.~(\ref{eq:BtoVformfactors}), the pre-factor of $V_1(q^2)$ is
$m_B+m_V$ while for $B\to A$ form factor $A_1(q^2)$, the factor
becomes $m_B-m_A$. We take $A_0$ and $T_1$ as an example to explain
the reason for the large $B\to A$ form factors. In
table~\ref{Tab:formfactorcomparison}, we give contributions from
three kinds of LCDAs to $T_1$: $\phi^T$, $\phi^v$ and $\phi^a$. The
contribution from $\phi^T$ is larger for $B\to a_1$, than the other
two transitions only because the axial-vector $a_1$ decay constant
is larger. Furthermore, larger axial vector meson mass induces
larger contributions from twist-3 distribution amplitudes $\phi^v$,
$\phi^a$ for both of $T_1^{B\to b_1}$ and $T_1^{B\to a_1}$.

\item Some $B\to A$ form factors strongly dependend on mixing
angles, which is obvious in Table
\ref{Tab:formfactorsBtoAaftermixing}. In our calculation for form
factors involving $f_1$ mesons, we have used the mixing angle
between the octet and singlet: $\theta=38^\circ(50^\circ)$ which is
very close to the ideal mixing angle $\theta=35.3^\circ$. That
implies the lighter meson $f_1(1285)$ is almost made up of
$\frac{\bar uu+\bar dd}{\sqrt 2}$ while the heavier meson
$f_1(1420)$ is dominated by the $\bar ss$ component. Thus $B\to
f_1(1420)$ and $B_s\to f_1(1285)$ form factors are suppressed by the
flavor structure and are numerically small. The form factors
involving $h_1$ are similar if the mixing angle is taken as
$45^\circ$.

\item The SU(3) symmetry breaking effect between $B\to a_1$ and
$B\to K_{1A}$ transition form factors is less than $10\%$. It is
also similar for the $B\to ^1P_1$ transition form factors.

\item In Table~\ref{Tab:formfactorsBtoAbeforemixing},
we can see that the form factor $A^{B\to K_{1A}}$ is almost equal to
$A^{B\to K_{1B}}$. But the physical states  $K_1(1270)$ and
$K_1(1400)$ are mixtures of $B\to K_{1A,1B}$. With the mixing angle
$\theta_K=\pm45^\circ$, the $B_{d,s}\to K_1(1270)(K_1(1400))$ form
factors are either enhanced by a factor $\sqrt 2$ or highly
suppressed.
\end{enumerate}

Up to now, there are many studies using some non-perturbative
methods on the $B\to A$ form factors: the constitute quark-meson
(CQM) model \cite{Deandrea:1998ww}, ISGW
\cite{Isgur:1988gb,Scora:1995ty}, QCD sum rules(QCDSR) and
light-cone sum rules(LCSR)
\cite{Lee:2006qj,Aliev:1999mx,Yang:BtoAinLCSR,Wang:2008bw} and
light-front quark model(LFQM) \cite{Cheng:2003sm,Cheng:2004yj}.
Results in LFQM and LCSR are collected in table
\ref{Tab:threeCompare} to make a comparison. These two approaches
are very different with the PQCD in the treatment of dynamics of
transition form factors, but at first we will analyze the
differences caused by non-perturbative inputs. For $B\to a_1$ and
$B\to K_{1A}$ form factors, most of our results (except $V_0$ and
$T_{1,2}$) are slightly larger than (or almost equal to) those
evaluated in LFQM, as slightly larger decay constants for $a_1$ and
$K_{1A}$ are used($f_{a_1}=203$ MeV and $f_{K_{1A}}=186$ MeV are
used by LFQM). Small differences in $V_0$ and $V_1$ have induced a
large difference in $V_2$, which could be reduced in future studies
using more precise hadronic inputs. As the decay constant of $b_1$
is zero in the isospin limit, the shape parameter $\omega$ in LFQM
can not be directly determined and the same value as that of $a_1$
is used\cite{Cheng:2003sm}. It is also similar for $K_{1B}$: they
used the same shape parameter as that of $K_{1A}$ which predicts
$f_{K_{1B}}=11$ MeV. Comparing with the QCDSR results
$f_{K_{1B}}=f_{K_{1B}}^T\times a_0^{||}$ given in
table~\ref{Table:Adecayconstant} and \ref{tab:AxialGegenbauer}, we
can see: although they are consistent within large theoretical
errors, the central value of $f_{K_{1B}}$ in QCDSR is larger than
the prediction in the LFQM. Thus our predictions for $B\to {^1P_1}$
form factors (central values) are larger than those in LFQM.
Compared with the recent LCSR results \cite{Yang:BtoAinLCSR}
collected in Table \ref{Tab:threeCompare}, our results differ from
theirs in two points:  one difference is that positive decay
constants for $1^1P_1$ mesons are adopted in LCSR, which leads to
the minus sign of the form factors for $1^1P_1$ mesons;  the other
difference is that form factor $A(0)$ in LCSR is larger than that in
the PQCD approach.

Experimentally,  the branching ratios of the color allowed
tree-dominated processes $B^0\to a_1^\pm \pi^\mp$ and $B^0\to
b_1^\pm \pi^\mp$ have been measured by the BaBar and Belle
collaborations~\cite{Aubert:2006dd,:2007jn,Aubert:2007xd} and
averaged by the heavy flavor averaging group~\cite{Barberio:2006bi}.
These two channels can be used to extract the $B\to a_1$ and $B\to
b_1$ form factors~\cite{Wang:2008hu}:
\begin{eqnarray}
 V_0^{B\to a_1}&=&(1.54\pm 0.28\pm0.03) f_+^{B\to \pi}=0.38 \pm 0.07 \pm
 0.01,\\
 V_0^{B\to b_1}&=&(1.45\pm 0.36\pm0.03) f_+^{B\to \pi}=0.35 \pm 0.03 \pm
 0.01,
\end{eqnarray}
where the penguin contributions are neglected for the small Wilson
coefficients. As we can see,  $V_0^{B\to a_1}$ is consistent with
our predictions within the errors, however $V_0^{B\to b_1}$ is
smaller than our predictions.

\subsection{Semilteptonic $B\to Al\bar\nu$ decays}
After integrating out the off shell W boson, one obtains the
effective Hamiltonian for $b\to ul\bar \nu_l$ transition
 \begin{eqnarray}
 {\cal H}_{eff}(b\to ul\bar \nu_l)=\frac{G_F}{\sqrt{2}}V_{ub}\bar
 u\gamma_{\mu}(1-\gamma_5)b \bar l\gamma^{\mu}(1-\gamma_5)\nu_l,
 \end{eqnarray}
where $V_{ub}$ is the CKM matrix element. With the form factors at
hand, the $\bar B\to Al\bar\nu_l$ decay widths are derived as:
\begin{eqnarray}
 \frac{d\Gamma_L(\bar B\to Al\bar\nu_l)}{dq^2}&=&(\frac{q^2-m_l^2}{q^2})^2\frac{ {\sqrt{\lambda(m_{B}^2,m_A^2,q^2)}}
  G_F^2 V_{ub}^2} {384m_{B}^3\pi^3}
 \times \frac{1}{q^2} \left\{ 3 m_l^2 \lambda(m_{B}^2,m_A^2,q^2) V_0^2(q^2)+\right.\nonumber\\
 &&\;\;\times  \left.(m_l^2+2q^2) \left|\frac{1}{2m_A}  \left[
 (m_{B}^2-m_A^2-q^2)(m_{B}-m_A)V_1(q^2)-\frac{\lambda(m_{B}^2,m_A^2,q^2)}{m_{B}-m_A}V_2(q^2)\right]\right|^2
 \right\},\\
 \frac{d\Gamma_\pm(\bar B\to Al\bar\nu_l)}{dq^2}&=&(\frac{q^2-m_l^2}{q^2})^2\frac{
 {\sqrt{\lambda(m_{B}^2,m_A^2,q^2)}} G_F^2V_{ub}^2}{384m_{B}^3\pi^3}
 \times   \nonumber\\
 &&\;\;\times \left\{ (m_l^2+2q^2) \lambda(m_{B}^2,m_A^2,q^2)\left|\frac{A(q^2)}{m_{B}-m_A}\mp
 \frac{(m_{B}-m_A)V_1(q^2)}{\sqrt{\lambda(m_{B}^2,m_A^2,q^2)}}\right|^2
 \right\},
\end{eqnarray}
where $\lambda(m_B^2,m_A^2,q^2)=(m_B^2+m_A^2-q^2)^2-4m_B^2m_A^2$,
$L$ and $\pm$ in the subscripts denote contributions from the
longitudinal polarization and the two transverse polarizations,
respectively. $m_l$ represents the mass of the charged lepton, and
$q^2$ is the momentum square of the lepton pair. Integrating over
the $q^2$, one obtains the longitudinal and transverse decay width
of $B\to Al\bar\nu$ decays.

For the semileptonic $B\to Al\bar\nu_l$ decays, physical quantities
$\rm{Br}_{\rm{L}}$, $\rm{Br}_{\rm{+}}$, $\rm{Br}_{\rm{-}}$,
$\rm{Br}_{\rm{total}}$ and
$\rm{Br}_{\rm{L}}/\rm{Br}_{\rm{T}}=\rm{\Gamma}_{\rm{L}}/\rm{\Gamma}_{\rm{T}}$
are predicted, where
$\rm{Br}_{\rm{T}}=\rm{Br}_{\rm{+}}+\rm{Br}_{\rm{-}}$ and
$\rm{Br}_{\rm{total}}=\rm{Br}_{\rm{L}}+\rm{Br}_{\rm{T}}$ with
$\rm{Br}_{\rm{L}}$, $\rm{Br}_{\rm{+}}$ and $\rm{Br}_{\rm{-}}$
corresponding to contributions of different polarizations to
branching ratios. Results for the $B\to A l\bar\nu_l$ ($l=e,\mu$)
and $B\to A \tau\bar\nu_{\tau}$ decays are listed in Table
\ref{tab:branchratios1} and \ref{tab:branchratios2} respectively,
with masses of the electron and muon neglected in the calculation.
There are some remarks:
 \begin{itemize}
\item Most of the branching ratios are of the order $10^{-4}$.
Some of the branching ratios are sensitive to the mixing angles,
especially for $K_1(1270)$ and $K_1(1400)$: one mixing angle gives
constructive contributions and the other gives destructive
contributions. Branching ratios for these two mixing angles are
two-order different, just as mentioned in the discussions about the
form factors. For the decays $B_s\to K_1(1270)$ and $B_s \to
K_1(1400)$, ratios of the contributions from longitudinal
polarization and transverse polarization are also much different
with each other.

\item The branching ratios of $B\to A
\tau\bar\nu_{\tau}$ decays are smaller than those of corresponding
$B\to A e\bar\nu_e$ decays, because the heavy $\tau$ lepton brings a
smaller phase space than the light electron.

\item Except for the $B_s \to K_1l\bar \nu_l$($l=e,\tau$) decay channels,
the ratios($\rm{Br}_{\rm{L}}/\rm{Br}_{\rm{T}}$) in $B \to
1^3P_1l\bar\nu_l$ decays are about $1.0\sim1.2$, while in $B\to
1^1P_1l\bar\nu_l$ decays, their values are roughly $2.0\sim2.5$. The
LCSR calculation\cite{Yang:BtoAinLCSR} has similar ratios for $B\to
1^3P_1l\bar\nu_l$ decays. However, their ratios for $B\to
1^1P_1l\bar\nu_l$ decays are around $0.5$. That means in these
decays the contributions of transverse polarization are relatively
larger in LCSR, which may be caused by their much larger form factor
$A$.
 \end{itemize}

%

%----------------------------------------------------------
%%%%%%%%%%%%%%%%%%%%%%%%%%%%%%%%%%%%%%%%%%%%%%%%%%%%%%%%%%%%%%%%%%%%%%%%%%%%%%%%%%%%%%%%%%%%%%%%%%%%
%%%%%%%%%%%     Results of Branch ratios in scenario 1(lnu) %%%%%%%%%%%%%%%%%%%%%%%%%%%%%%%%%%%%%%%%
%%%%%%%%%%%%%%%%%%%%%%%%%%%%%%%%%%%%%%%%%%%%%%%%%%%%%%%%%%%%%%%%%%%%%%%%%%%%%%%%%%%%%%%%%%%%%%%%%%%%
 \begin{table}
 \caption{The total branching ratios for the $b\to ul\bar
 \nu_l$ ($l=e,\mu$ and unit $10^{-4}$ for branching ratios). $\rm{Br}_{\rm{L}}$ and $\rm{Br}_{\pm}$ are
the longitudinally and transversely polarized contributions to the
branching ratios. And $\rm{Br}_{\rm{T}}=\rm{Br}_{+} + \rm{Br}_{-}$.
For the decays with a mixing meson, results in the first lines are
calculated using $\theta_K=45^\circ$, $\theta_{^1P_1}=10^\circ$ or
$\theta_{^3P_1}=38^\circ$, while the second line corresponds to the
angle $\theta_K=-45^\circ$, $\theta_{^1P_1}=45^\circ$ or
$\theta_{^3P_1}=50^\circ$.}
 \label{tab:branchratios1}
 \begin{center}
 \begin{tabular}{|c|ccccc|}
 \hline\hline
 \ \ \      & $\rm{Br}_{\rm{L}}$  &$\rm{Br}_+$  &$\rm{Br}_-$   &$\rm{Br}_{\rm{total}}$  &$\rm{Br}_{\rm{L}}/\rm{Br}_{\rm{T}}$ \\
 \hline
 \ \ \ $\bar B^0\to a_1^+$         &$1.60_{-0.82}^{+1.03}$    &$0.04_{-0.02}^{+0.02}$  &$1.31_{-0.56}^{+0.71}$  &$2.96_{-1.39}^{+1.74}$  &$1.18_{-0.22}^{+0.20}$  \\
 \hline
 \ \ \ $\bar B^0\to b_1^+$         &$2.10_{-0.86}^{+1.07}$    &$0.03_{-0.01}^{+0.02}$  &$0.76_{-0.36}^{+0.44}$  &$2.88_{-1.22}^{+1.51}$  &$2.67_{-0.35}^{+0.44}$   \\
 \hline
 \ \ \ $B^-\to f_1^0(1285)$   &$0.93_{-0.48}^{+0.60}$    &$0.03_{-0.01}^{+0.01}$  &$0.73_{-0.31}^{+0.39}$  &$1.69_{-0.79}^{+0.99}$  &$1.22_{-0.23}^{+0.21}$   \\
 \ \ \                        &$0.87_{-0.44}^{+0.55}$    &$0.02_{-0.01}^{+0.01}$    &$0.69_{-0.29}^{+0.37}$    &$1.58_{-0.74}^{+0.92}$    &$1.22_{-0.23}^{+0.22}$  \\
 \hline
 \ \ \ $B^-\to f_1^0(1420)$   &$\sim0.002$    &$<0.001$   &$\sim0.001$  &$\sim 0.003$  &$1.12_{-0.63}^{+0.47}$   \\
 \ \ \                        &$0.05_{-0.03}^{+0.04}$    &$\sim 0.001$    &$0.04_{-0.02}^{+0.02}$    &$0.09_{-0.05}^{+0.06}$    &$1.25_{-0.31}^{+0.27}$  \\
 \hline
 \ \ \ $B^-\to h_1^0(1170)$   &$1.08_{-0.44}^{+0.55}$    &$0.02_{-0.01}^{+0.01}$  &$0.43_{-0.20}^{+0.25}$  &$1.53_{-0.65}^{+0.80}$  &$2.41_{-0.31}^{+0.37}$   \\
 \ \ \                        &$1.38_{-0.56}^{+0.70}$    &$0.02_{-0.01}^{+0.01}$    &$0.54_{-0.26}^{+0.31}$    &$1.94_{-0.82}^{+1.02}$    &$2.42_{-0.32}^{+0.39}$  \\
 \hline
 \ \ \ $B^-\to h_1^0(1380)$   &$0.23_{-0.10}^{+0.12}$    &$\sim 0.004$  &$0.09_{-0.04}^{+0.05}$  &$0.32_{-0.14}^{+0.17}$  &$2.64_{-0.35}^{+0.45}$   \\
 \ \ \                        &$0.01_{-0.01}^{+0.01}$    &$<0.001$                &$0.01_{-0.0.00}^{+0.00}$    &$0.02_{-0.01}^{+0.01}$    &$2.45_{-0.30}^{+0.32}$  \\
 \hline\hline
 \ \ \ $\bar B_s\to K_1^+(1270)$ &$3.65_{-1.87}^{+2.27}$    &$0.08_{-0.04}^{+0.05}$  &$2.01_{-1.00}^{+1.21}$  &$5.75_{-2.89}^{+3.49}$  &$1.74_{-0.30}^{+0.30}$  \\
 \ \ \                        &$0.01_{-0.00}^{+0.04}$    &$< 0.001$    &$0.03_{-0.03}^{+0.05}$    &$0.04_{-0.02}^{+0.06}$    &$0.16_{-0.15}^{+4.00}$  \\
 \hline
 \ \ \ $\bar B_s\to K_1^+(1400)$ &$\sim0.005$    &$< 0.001$  &$0.03_{-0.03}^{+0.04}$  &$0.03_{-0.02}^{+0.05}$  &$0.16_{-0.15}^{+4.05}$  \\
 \ \ \                        &$3.00_{-1.54}^{+1.87}$    &$0.06_{-0.03}^{+0.03}$    &$1.59_{-0.79}^{+0.96}$    &$4.65_{-2.34}^{+2.82}$    &$1.83_{-0.31}^{+0.31}$  \\
 \hline\hline
 \end{tabular}
 \end{center}
 \end{table}
%----------------------------------------------------------
%%%%%%%%%%%%%%%%%%%%%%%%%%%%%%%%%%%%%%%%%%%%%%%%%%%%%%%%%%%%%%%%%%%%%%%%%%%%%%%%%%%%%%%%%%%%%%%%%%%%
%%%%%%%%%%%     Results of Branch ratios in scenario 1(lnu) %%%%%%%%%%%%%%%%%%%%%%%%%%%%%%%%%%%%%%%%
%%%%%%%%%%%%%%%%%%%%%%%%%%%%%%%%%%%%%%%%%%%%%%%%%%%%%%%%%%%%%%%%%%%%%%%%%%%%%%%%%%%%%%%%%%%%%%%%%%%%
 \begin{table}
 \caption{The same as Table \ref{tab:branchratios1} except $b\to u\tau\bar \nu_{\tau}$.}
 \label{tab:branchratios2}
 \begin{center}
 \begin{tabular}{|c|ccccc|}
 \hline\hline
 \ \ \      & $\rm{Br}_{\rm{L}}$  &$\rm{Br}_+$  &$\rm{Br}_-$   &$\rm{Br}_{\rm{total}}$  &$\rm{Br}_{\rm{L}}/\rm{Br}_{\rm{T}}$ \\
 \hline
 \ \ \ $\bar B^0\to a_1^+$         &$0.69_{-0.36}^{+0.44}$    &$0.03_{-0.01}^{+0.01}$  &$0.62_{-0.27}^{+0.33}$  &$1.34_{-0.63}^{+0.78}$  &$1.06_{-0.20}^{+0.18}$  \\
 \hline
 \ \ \ $\bar B^0\to b_1^+$         &$0.88_{-0.36}^{+0.45}$    &$0.02_{-0.01}^{+0.01}$  &$0.36_{-0.17}^{+0.21}$  &$1.26_{-0.54}^{+0.66}$  &$2.32_{-0.29}^{+0.38}$   \\
 \hline
 \ \ \ $B^-\to f_1^0(1285)$   &$0.39_{-0.20}^{+0.25}$    &$0.02_{-0.01}^{+0.01}$  &$0.34_{-0.14}^{+0.18}$  &$0.74_{-0.35}^{+0.43}$  &$1.08_{-0.21}^{+0.19}$   \\
 \ \ \                        &$0.36_{-0.19}^{+0.23}$    &$0.02_{-0.01}^{+0.01}$  &$0.32_{-0.13}^{+0.17}$  &$0.70_{-0.32}^{+0.40}$  &$1.09_{-0.21}^{+0.19}$   \\
 \hline
 \ \ \ $B^-\to f_1^0(1420)$   &$\sim 0.001$    &$<0.001$          &$\sim 0.001$  &$\sim 0.001$  &$0.96_{-0.53}^{+0.40}$   \\
 \ \ \                        &$0.02_{-0.01}^{+0.01}$    &$\sim 0.001$  &$0.02_{-0.01}^{+0.01}$  &$0.03_{-0.02}^{+0.02}$  &$1.08_{-0.27}^{+0.23}$   \\
 \hline
 \ \ \ $B^-\to h_1^0(1170)$   &$0.47_{-0.19}^{+0.24}$    &$0.01_{-0.01}^{+0.01}$  &$0.21_{-0.10}^{+0.12}$  &$0.70_{-0.30}^{+0.37}$  &$2.12_{-0.27}^{+0.33}$   \\
 \ \ \                        &$0.60_{-0.25}^{+0.31}$    &$0.02_{-0.01}^{+0.01}$  &$0.27_{-0.13}^{+0.15}$  &$0.89_{-0.38}^{+0.47}$  &$2.13_{-0.27}^{+0.34}$   \\
 \hline
 \ \ \ $B^-\to h_1^0(1380)$   &$0.09_{-0.04}^{+0.05}$    &$\sim 0.002$  &$0.04_{-0.02}^{+0.02}$  &$0.13_{-0.05}^{+0.07}$  &$2.23_{-0.29}^{+0.38}$   \\
 \ \ \                        &$0.01_{-0.00}^{+0.00}$    &$<0.001$                  &$\sim 0.002$  &$0.01_{-0.00}^{+0.00}$  &$2.09_{-0.25}^{+0.26}$   \\
 \hline \hline
 \ \ \ $\bar B_s\to K_1^+(1270)$     &$1.59_{-0.81}^{+0.98}$    &$0.05_{-0.03}^{+0.03}$  &$0.98_{-0.49}^{+0.59}$  &$2.62_{-1.31}^{+1.58}$  &$1.54_{-0.27}^{+0.27}$  \\
 \ \ \                           &$\sim 0.003$     &$<0.001$  &$0.01_{-0.01}^{+0.02}$  &$0.02_{-0.01}^{+0.03}$  &$0.17_{-0.12}^{+3.58}$   \\
 \hline
 \ \ \ $\bar B_s\to K_1^+(1400)$     &$\sim 0.002$    &$<0.001$  &$0.01_{-0.01}^{+0.02}$  &$0.01_{-0.01}^{+0.02}$  &$0.18_{-0.12}^{+3.57}$  \\
 \ \ \                        &$1.18_{-0.60}^{+0.73}$    &$0.04_{-0.02}^{+0.02}$  &$0.71_{-0.36}^{+0.43}$  &$1.93_{-0.97}^{+1.17}$  &$1.58_{-0.27}^{+0.27}$   \\
 \hline\hline
 \end{tabular}
 \end{center}
 \end{table}

\subsection{More Discussions on the Mixing between $K_{1A}$ and $K_{1B}$}

As pointed out, the $B\to K_1(1270)$ and $B\to K_1(1400)$ form
factors have either quite large or quite small values, for the
mixing angles are $\pm 45^\circ$.  Actually, these two values are
just chosen for illustration, as the determination in $\tau$ decays
are not stringent. There are some attempts to determine the mixing
angles between the two $K_{1}$ mesons in $B$ meson decays. For
example, the authors in Ref. \cite{Cheng:2003sm} found that the
mixing angle between $K_{1A}$ and $K_{1B}$ is two-fold:
$\theta=38^\circ$ or $\theta=50^\circ$. However, their determination
depends on the LFQM predictions on the $B\to A$ form factors, which
is model-dependent. To reduce the uncertainties caused by the
dynamics of strong interactions, we propose to use the $\bar B^0\to
D^+ K_{1}^-$ decay to extract the mixing angle between these two
mesons. The dynamics of this charmful decay is very similar to that
of $\bar B^0\to D^+\pi^-$. Neglecting the higher power corrections,
the decay amplitudes of $\bar B^0\to D^+ M^-$ ($M^-$ denotes $\pi^-$
or $K_1^-$) can be factorized into the $B\to D$ form factor and a
convolution of a hard kernel with the light-cone distribution
amplitude of the emitted light meson~\cite{Bauer:2001cu}. To the
leading order in $\alpha_s$, the convolution reduces to the decay
constant of the emitted light meson. Then the factorization formula
is proved to have the form:
\begin{eqnarray}
 {\cal A}(\bar B^0\to D^+ M^-)&=& \frac{G_F}{\sqrt 2}m_B^2 f_M
 F_0^{B\to D}(1-r^2)^3 V_{cb}V_{ud(s)}^* a_1,
\end{eqnarray}
where $r=m_D/m_{B}$. Due to the small value for the longitudinal
decay constant $f_{K_{1B}}$, the decay amplitude of $\bar B^0\to
D^+K_{1B}$ is very small. Thus the physical decay channels receive
the leading contributions from $\bar B^0\to D^+ K_{1A}$. Utilizing
the $B\to D$ form factors which are well explored in the heavy quark
effctive theory, we can directly present our predictions on $\bar
B^0\to D^+ K_{1}$ decays, if the mixing angle is known. On the other
hand, one can also obtain the mixing angle, if the experimental data
on the branching ratio is provided. In practice, in order to reduce
the uncertainty from the nonperturbative inputs, one can use the
experimental data of the branching fraction of $\bar B^0\to
D^+\pi^-$ instead of any theoretical model. The ratios of branching
fractions are given as
\begin{eqnarray}
 R\equiv \frac{ {\cal BR}(\bar B^0\to D^+ K_1^-)}{ {\cal BR}(\bar B^0\to
 D^+\pi^-)}&=&\frac{f_{K_1}^2 |V_{us}|^2}{f_{\pi}^2 |V_{ud}|^2},
\end{eqnarray}
where $f_{K_1}$ is the decay constant for a physical state.

\begin{figure}%[]
\begin{center}
\includegraphics[width=7cm]{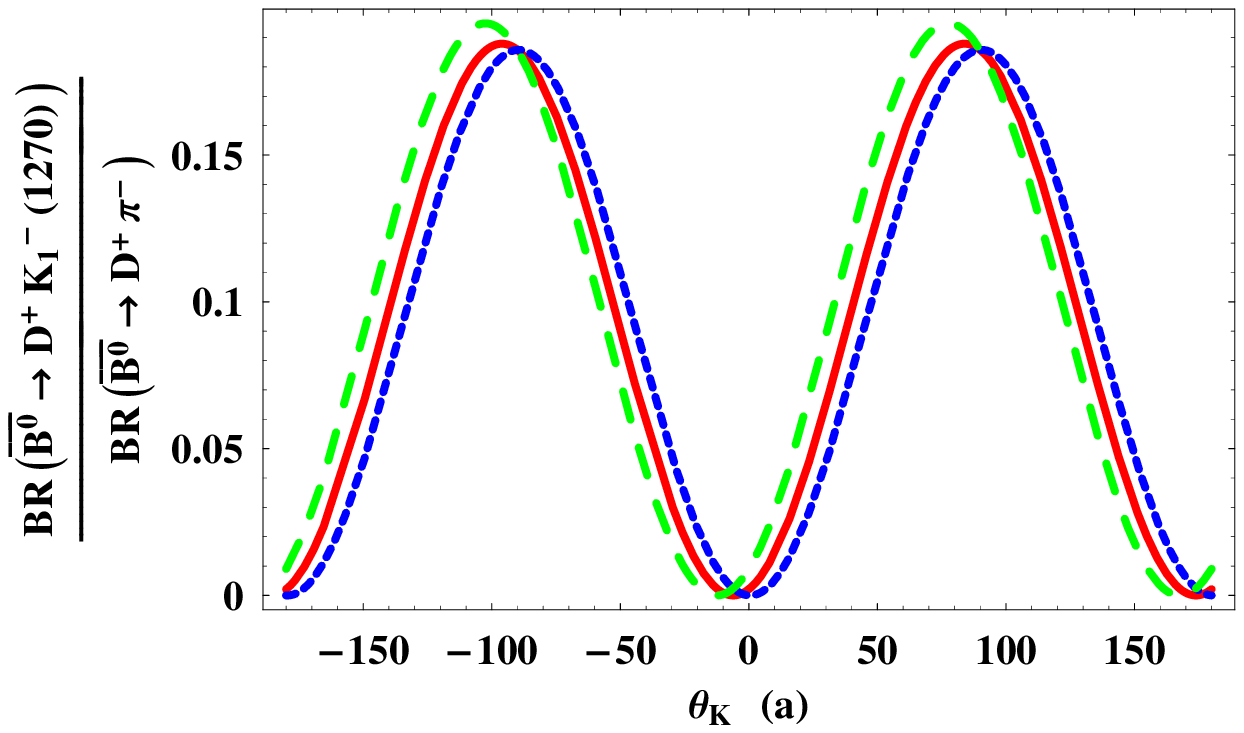}
\hspace{3mm}
\includegraphics[width=7cm]{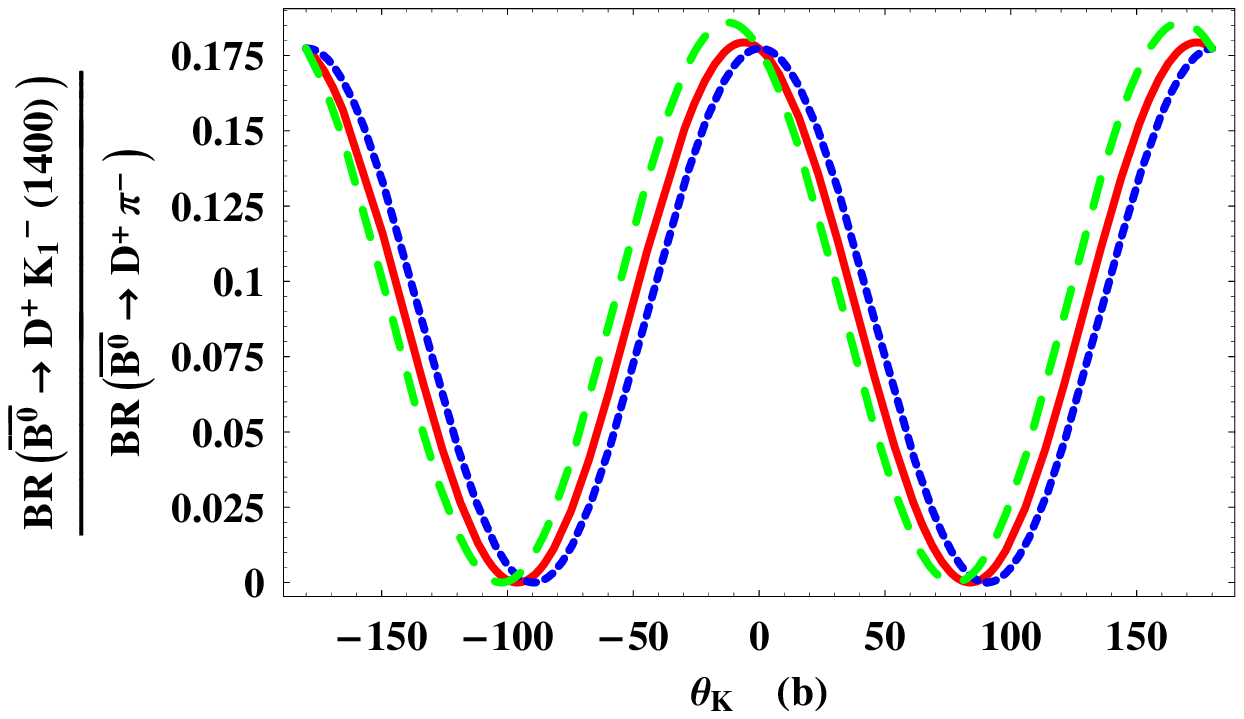}
\end{center}
\caption{The dependence of  ratios of branching fractions on the
mixing angle $\theta_K$ as discussed in the text. The left diagram
(a) denotes the ratio  $\frac{ {\cal BR}(\bar B^0\to D^+
K_1^-(1270))}{ {\cal BR}(\bar B^0\to  D^+\pi^-)}$, while the right
diagram (b) denotes the ratio  $\frac{ {\cal BR}(\bar B^0\to D^+
K_1^-(1400))}{ {\cal BR}(\bar B^0\to  D^+\pi^-)}$. The uncertainties
caused by the Gegenbauer moment $a_{0}^{|| K_{1B}}$ are shown in
these diagrams: the red solid line denotes the central value, while
the blue short-dashed (green long-dashed) line denotes the lower
(upper) uncertainty.  } \label{Feyn:mixing}
\end{figure}

In the ratios for the two channels $\bar B\to DK_1(1270)$ and $\bar
B\to DK_1(1400)$, the main uncertainties come from the decay
constants of $K_1(1270)$ and $K_1(1400)$ which are combinations of
the two decay constants $f_{K_{1A}}$ and
$f_{K_{1B}}=f_{K_{1B}}\times a_0^{\parallel K_{1B}}$. From the
table~\ref{Table:Adecayconstant} and  \ref{tab:AxialGegenbauer}, we
can see the parameter $a_0^{\parallel K_{1B}}$ has the largest
uncertainty.  In Fig.\ref{Feyn:mixing}, we plot the dependence on
the mixing angle of the branching ratios utilizing the decay
constants evaluated in the sum rules and we also take the
uncertainty of $a_0^{\parallel K_{1B}}$ into account for the error
estimation:  the left diagram (a) denotes the ratio  $\frac{ {\cal
BR}(\bar B^0\to D^+ K_1^-(1270))}{ {\cal BR}(\bar B^0\to
D^+\pi^-)}$, while the right diagram (b) denotes the ratio  the
ratio  $\frac{ {\cal BR}(\bar B^0\to D^+ K_1^-(1400))}{ {\cal
BR}(\bar B^0\to  D^+\pi^-)}$. The uncertainties caused by the
Gegenbauer moment $a_{0}^{|| K_{1B}}$ are shown in these diagrams:
the red solid line denotes the central value, while the blue
short-dashed (green long-dashed) line denotes the lower (upper)
uncertainty.  Once the experimental data are available in the
future, these two diagrams can be used to extract the mixing angles
in model-independent way. As an illustration, we will give our
predictions utilizing the decay constants extracted from the $\tau$
decays. The branching ratio of $\bar B^0\to D^+\pi^-$ has been
averaged as~\cite{Barberio:2006bi}:
\begin{eqnarray}
 {\cal BR}(\bar B^0\to D^+\pi^-)&=&(2.65\pm0.15)\times
 10^{-3},
\end{eqnarray}
which gives the following predictions on the branching fractions:
\begin{eqnarray}
 {\cal BR}(\bar B^0\to D^+K_1^-(1270))&=&(2.1\pm0.5)\times
 10^{-4},\\
 {\cal BR}(\bar B^0\to D^+K_1^-(1400))&=&(1.2^{+1.8}_{-1.2})\times
 10^{-4}.
\end{eqnarray}
These results will be certainly tested on the future experiments and
the measurements are  very helpful to detect the internal structure
of $K_1(1270)$ and  $K_1(1400)$.

%%%%%%%%%%%%%%%%%%%%%%%%%%%%%%%%%%%%%%%%%%%%%%%%%%%%%%%%%%
\section{Summary}
%%%%%%%%%%%%%%%%%%%%%%%%%%%%%%%%%%%%%%%%%%%%%%%%%%%%%%%%%%

The PQCD approach is based on $k_T$ factorization where we keep the
transverse momentum of valence quarks in the mesons to smear the
endpoint singularity. $k_T$ resummation of double logarithms results
in the Sudakov factor. Resummation of double logarithms from the
threshold region leads to the jet function. Sudakov factor and jet
function can suppress the contribution from the large $b$ region and
small $x$ region, respectively. This makes the PQCD approach
self-consistent. Inspired by the success of the PQCD approach in
non-leptonic B decays \cite{pqcd}, we give a comprehensive study on
the charmless $B\to A$ transition form factors and the semileptonic
$B\to Al\bar\nu$ decays in the PQCD approach.

Semi-leptonic and radiative decays are somewhat simpler than
non-leptonic decays as only one hadronic meson involved in the final
state. In this case, the dominant amplitude can be parameterized
into form factors. In order to make precise prediction and extract
the CKM matrix elements, we have to know the behavior of form
factors. In the PQCD approach, the final state meson moves nearly on
the light-cone and a hard-gluon-exchange is required. Thus the
dominant contribution is from the hard region which can be
factorized. In section \ref{section:formfactor}, we have used the
same input hadronic parameters with Ref. \cite{Ali:2007ff} and
updated all the $B\to V$ decay form factors in the PQCD approach.
Compared with the results evaluated from other approaches, we find
that despite a number of theoretical differences in different
approaches, all the numerical results of the form factors are
surprisingly consistent with each other.

In section~\ref{section:formfactor}, we study $B\to A$ form factors.
As the quark contents for the axial-vectors have not been uniquely
determined, we give two different sets of results for the form
factors according to different mixing angles. For the axial-vector
mesons $f_1$, we have used the mixing angle between the octet and
singlet: $\theta=38^\circ(50^\circ)$ which is close to the ideal
mixing angle $\theta=35.3^\circ$. With this mixing angle, one can
easily check that the lighter meson $f_1(1285)$ is made almost up of
$\frac{\bar uu+\bar dd}{\sqrt 2}$ while the heavier meson
$f_1(1420)$ is composed of $\bar ss$. Thus partial decay widths of
$B\to f_1(1420)l\bar\nu$ and $B_s\to f_1(1285)l\bar\nu$ are
suppressed by the flavor structure.

The mixing angle between the two strange mesons $K_1(1270)$ and
$K_1(1400)$ has large ambiguities. In order to reduce these
ambiguities,  we propose to use the $\bar B^0\to D^+ K_{1}^-$ decay
to extract the mixing angle between these two mesons. Our method is
model-independent which receives very small uncertainties.  In
Fig.~\ref{Feyn:mixing}, we show the strong dependence of the $\bar
B^0\to D^+ K_{1}$ decay branching ratio on the mixing angle
$\theta_K$.  Our calculation can be used to constrain this mixing
angle using experimental measurements. These studies of higher
resonance production in $B$ decays can help us to uncover the
mysterious structure of these excited states.

%%%%%%%%%%%%%%%%%%%%%%%%%%%%%%%%%%%%%%%%%%%%%%%%%%%%%%%%%%
\section*{Acknowledgements}
%%%%%%%%%%%%%%%%%%%%%%%%%%%%%%%%%%%%%%%%%%%%%%%%%%%%%%%%%%

This work is partly supported by National Natural Science Foundation
of China under the Grant No. 10735080, 10625525, 10525523, and
10805037, and partly supported by Project of Knowledge Innovation
Program (PKIP) of Chinese Academy of Sciences, under Grant No.
KJCX2.YW.W10. We would like to acknowledge S.-Y. Li, Y. Li, Y.-L.
Shen, X.-X. Wang, Y.-M. Wang, Z.-T. Wei, K.-C. Yang, M.-Z. Yang and
H. Zou for valuable discussions.

%%%%%%%%%%%%%%%%%%%%%%%%%%%%%%%%%%%%%%%%%%%%%%%%%%%%%%%%%%
\appendix
%%%%%%%%%%%%%%%%%%%%%%%%%%%%%%%%%%%%%%%%%%%%%%%%%%%%%%%%%%

%%%%%%%%%%%%%%%%%%%%%%%%%%%%%%%%%%%%%%%%%%%%%%%%%%%%%%%%%%
\section{pQCD functions}\label{PQCDfunctions}
%%%%%%%%%%%%%%%%%%%%%%%%%%%%%%%%%%%%%%%%%%%%%%%%%%%%%%%%%%

In this appendix, we group the functions which appear in the
factorization formulae. The hard scales are chosen as
\begin{eqnarray}
 t_e^1&=&\mbox{max}\{{\sqrt {x_2}m_{B}t_c,1/b_1,1/b_2}\},\;\;\;
 t_e^2=\mbox{max}\{{\sqrt{x_1}m_{B}t_c,1/b_1,1/b_2}\},
\end{eqnarray}
where $t_c$ is a factor varying from $0.75$ to $1.25$ for error
estimations.

%\begin{eqnarray}
%&&\int d^2b_1d^2b_2\int
%\frac{d^2k_1}{(2\pi)^2}\frac{d^2k_2}{(2\pi)^2}
%\frac{1}{\alpha^2+k_1^2}\frac{1}{\beta^2+(k_1+k_2)^2}e^{-ik_1\cdot
%b_1}e^{-ik_2\cdot b_2}\nonumber\\
%&=&\int^\infty_0b_1db_1\int^\infty_0b_2db_2K_0(\beta
%b_2)[\theta(b_2-b_1)I_0(\alpha b_1)K_0(\alpha
%b_2)+\theta(b_1-b_2)I_0(\alpha b_2)K_0(\alpha b_1)]
%\end{eqnarray}
The functions $h_i$ in decay amplitudes are from the propagators of
virtual quark and gluon and are defined by:
\begin{eqnarray}
h_e(A,B,b_1,b_2)&=&\Big[\theta(A)K_0(\sqrt
{A}m_Bb_1)+\theta(-A)i\frac{\pi}{2}H_0(\sqrt
{-A}m_{B}b_1)\Big]\nonumber\\
&&\times \bigg\{\theta(b_1-b_2)\Big[\theta(B)K_0(\sqrt B
m_{B}b_1)I_0(\sqrt Bm_{B}b_2)\nonumber\\
&&\;\;\;+\theta(-B)i\frac{\pi}{2}H_0^{(1)}(\sqrt {-B}
m_{B}b_1)J_0(\sqrt {-B}m_{B}b_2)\Big]+ (b_1\leftrightarrow
b_2)\bigg\},
\end{eqnarray}
where $H_0^{(1)}(z) = \mathrm{J}_0(z) + i\, \mathrm{Y}_0(z)$.
% (A isthe virtuality of the gluon and B is the virtuality of the quark
%propagator, $b_2$ is the coordinate for the quark propagator).

The Sudakov factor from threshold resummation is universal,
independent of flavors of internal quarks, twists,  and the specific
processes. To simplify the analysis, the following parametrization
has been used  \cite{Li:2002mi}:
\begin{eqnarray}
S_t(x)=\frac{2^{1+2c}\Gamma(3/2+c)}{\sqrt{\pi}\Gamma(1+c)}
[x(1-x)]^c\;, \label{str}
\end{eqnarray}
with $c=0.4\pm0.1$. This parametrization, symmetric under the
interchange of $x$ and $1-x$, is convenient for evaluation of the
amplitudes. It is obvious that the threshold resummation modifies
the end-point behavior of the meson distribution amplitudes,
rendering them vanish at $x\to 0$ or $1$.

Function $S_{ab}(t)$ in Sudakov factors is given by
\begin{eqnarray}
S_{ab}(t)=S_B(t)+S_2(t)],
\end{eqnarray}
in which $S_B(t)$ and $S_2(t)$ are defined as
\begin{eqnarray}
S_B(t)&=&s\left(x_1\frac{m_{B}}{\sqrt
2},b_1\right)+\frac{5}{3}\int^t_{1/b_1}\frac{d\bar \mu}{\bar
\mu}\gamma_q(\alpha_s(\bar \mu)),\\
S_2(t)&=&s\left(x_2\frac{m_{B}}{\sqrt
2},b_2\right)+s\left((1-x_2)\frac{m_{B}}{\sqrt
2},b_2\right)+2\int^t_{1/b_2}\frac{d\bar \mu}{\bar
\mu}\gamma_q(\alpha_s(\bar \mu)),
\end{eqnarray}
with the quark anomalous dimension $\gamma_q=-\alpha_s/\pi$. The
explicit form for the  function $s(Q,b)$ is:
\begin{eqnarray}
s(Q,b)&=&~~\frac{A^{(1)}}{2\beta_{1}}\hat{q}\ln\left(\frac{\hat{q}}
{\hat{b}}\right)-
\frac{A^{(1)}}{2\beta_{1}}\left(\hat{q}-\hat{b}\right)+
\frac{A^{(2)}}{4\beta_{1}^{2}}\left(\frac{\hat{q}}{\hat{b}}-1\right)
%\nonumber \\
-\left[\frac{A^{(2)}}{4\beta_{1}^{2}}-\frac{A^{(1)}}{4\beta_{1}}
\ln\left(\frac{e^{2\gamma_E-1}}{2}\right)\right]
\ln\left(\frac{\hat{q}}{\hat{b}}\right)
\nonumber \\
&&+\frac{A^{(1)}\beta_{2}}{4\beta_{1}^{3}}\hat{q}\left[
\frac{\ln(2\hat{q})+1}{\hat{q}}-\frac{\ln(2\hat{b})+1}{\hat{b}}\right]
+\frac{A^{(1)}\beta_{2}}{8\beta_{1}^{3}}\left[
\ln^{2}(2\hat{q})-\ln^{2}(2\hat{b})\right],
%\nonumber \\
%&&+\frac{A^{(1)}\beta_{2}}{8\beta_{1}^{3}}
%\ln\left(\frac{e^{2\gamma_E-1}}{2}\right)\left[
%\frac{\ln(2\hat{q})+1}{\hat{q}}-\frac{\ln(2\hat{b})+1}{\hat{b}}\right]
%-\frac{A^{(2)}\beta_{2}}{16\beta_{1}^{4}}\left[
%\frac{2\ln(2\hat{q})+3}{\hat{q}}-\frac{2\ln(2\hat{b})+3}{\hat{b}}\right]
%\nonumber \\
%& &-\frac{A^{(2)}\beta_{2}}{16\beta_{1}^{4}}
%\frac{\hat{q}-\hat{b}}{\hat{b}^2}\left[2\ln(2\hat{b})+1\right]
%+\frac{A^{(2)}\beta_{2}^2}{432\beta_{1}^{6}}
%\frac{\hat{q}-\hat{b}}{\hat{b}^3}
%\left[9\ln^2(2\hat{b})+6\ln(2\hat{b})+2\right]
%\nonumber \\
%&& +\frac{A^{(2)}\beta_{2}^2}{1728\beta_{1}^{6}}\left[
%\frac{18\ln^2(2\hat{q})+30\ln(2\hat{q})+19}{\hat{q}^2}
%-\frac{18\ln^2(2\hat{b})+30\ln(2\hat{b})+19}{\hat{b}^2}\right],
\end{eqnarray} where the variables are defined by
\begin{eqnarray}
\hat q\equiv \mbox{ln}[Q/(\sqrt 2\Lambda)],~~~ \hat b\equiv
\mbox{ln}[1/(b\Lambda)], \end{eqnarray} and the coefficients
$A^{(i)}$ and $\beta_i$ are \begin{eqnarray}
\beta_1=\frac{33-2n_f}{12},~~\beta_2=\frac{153-19n_f}{24},\nonumber\\
A^{(1)}=\frac{4}{3},~~A^{(2)}=\frac{67}{9}
-\frac{\pi^2}{3}-\frac{10}{27}n_f+\frac{8}{3}\beta_1\mbox{ln}(\frac{1}{2}e^{\gamma_E}),
\end{eqnarray}
$n_f$ is the number of the quark flavors and $\gamma_E$ is the Euler
constant. We will use the one-loop running coupling constant, i.e.
we pick up only the four terms in the first line of the expression
for the function $s(Q,b)$.

%\newpage
%%%%%%%%%%%%%%%%%%%%%%%%%%%%%%%%%%%%%%%%%%%%%%%%%%%%%%%%%%

%%%%%%%%%%%%%%%%%%%%%%%%%%%%%%%%%%%%%%%%%%%%%%%%%%%%%%%%%%%%

\begin{thebibliography}{11}
%


%\cite{Keum:2000ph}
\bibitem{Keum:2000ph}
  Y.~Y.~Keum, H.~n.~Li and A.~I.~Sanda,
  %``Fat penguins and imaginary penguins in perturbative QCD,''
  Phys.\ Lett.\  B {\bf 504}, 6 (2001)
  [arXiv:hep-ph/0004004];
  %%CITATION = PHLTA,B504,6;%%
  %``Penguin enhancement and B --> K pi decays in perturbative QCD,''
  Phys.\ Rev.\  D {\bf 63}, 054008 (2001)
  [arXiv:hep-ph/0004173];
  %%CITATION = PHRVA,D63,054008;%%

  C.~D.~Lu, K.~Ukai and M.~Z.~Yang,
  %``Branching ratio and CP violation of B --> pi pi decays in perturbative  QCD
  %approach,''
  Phys.\ Rev.\  D {\bf 63}, 074009 (2001)
  [arXiv:hep-ph/0004213].
  %%CITATION = PHRVA,D63,074009;%%

\bibitem{direct}
B.~H.~Hong and C.~D.~Lu,
  %``Direct CP violation in hadronic B decays,''
  Sci.\ China {\bf G49}, 357 (2006)
  [arXiv:hep-ph/0505020].
  %%CITATION = SCASE,G49,357;%%


\bibitem{kphi}
H.-n Li, and S. Mishima, Phys. Rev. D{\bf71}, 054025 (2005)
[hep-ph/0411146]; H.-n. Li, Phys. Lett. B{\bf 622}, 63 (2005)
[hep-ph/0411305].

\bibitem{kphi-e}
A.V. Gritsan,  Invited talk at 5th Flavor Physics and CP Violation
Conference (FPCP 2007), Bled, Slovenia, 12-16 May 2007;
arXiv:0706.2030 [hep-ex]


%\cite{Abe:2001wa}
\bibitem{Abe:2001wa}
  K.~Abe {\it et al.}  [Belle Collaboration],
  %``Observation of B --> J/psi K1(1270),''
  Phys.\ Rev.\ Lett.\  {\bf 87}, 161601 (2001)
  [arXiv:hep-ex/0105014].
  %%CITATION = PRLTA,87,161601;%%

%\cite{Aubert:2002sp}
\bibitem{Aubert:2002sp}
  B.~Aubert {\it et al.}  [BABAR Collaboration],
  %``Measurement of the B0 --> D*- a1+ branching fraction with partially
  %reconstructed D*,''
  arXiv:hep-ex/0207085.
  %%CITATION = HEP-EX/0207085;%%




%\cite{Barberio:2006bi}
\bibitem{Barberio:2006bi}
  E.~Barberio {\it et al.}  [Heavy Flavor Averaging Group (HFAG)],
  %``Averages of b-hadron properties at the end of 2005,''
  arXiv:hep-ex/0603003.
  %%CITATION = HEP-EX/0603003;%%
%


%\cite{Yang:2005gk}
\bibitem{Yang:2005gk}
  K.~C.~Yang,
  %``Light-cone distribution amplitudes for the light 1(1)P(1) mesons,''
  JHEP {\bf 0510}, 108 (2005)
  [arXiv:hep-ph/0509337].
  %%CITATION = JHEPA,0510,108;%%

%\cite{Yang:2007zt}
\bibitem{Yang:2007zt}
  K.~C.~Yang,
  %``Light-cone distribution amplitudes of axial-vector mesons,''
  Nucl.\ Phys.\  B {\bf 776}, 187 (2007)
  [arXiv:0705.0692 [hep-ph]].
  %%CITATION = NUPHA,B776,187;%%



%\cite{Wang:2007an}
\bibitem{Wang:2007an}
  W.~Wang, R.~H.~Li and C.~D.~Lu,
  %``Radiative charmless B_{(s)}\to V \gamma and B_{(s)}\to A \gamma decays in
  %pQCD approach,''
  arXiv:0711.0432 [hep-ph]. The manuscript is being revised.
  %%CITATION = ARXIV:0711.0432;%%


%\cite{Lu:2002ny}
\bibitem{Lu:2002ny}
  T.~Kurimoto, H.~n.~Li and A.~I.~Sanda,
  %``Leading power contributions to B --> pi, rho transition form factors,''
  Phys.\ Rev.\  D {\bf 65}, 014007 (2002)
  [arXiv:hep-ph/0105003];
  %%CITATION = PHRVA,D65,014007;%%;
  C.~D.~Lu and M.~Z.~Yang,
  %``B to light meson transition form factors calculated in perturbative QCD
  %approach,''
  Eur.\ Phys.\ J.\  C {\bf 28}, 515 (2003)
  [arXiv:hep-ph/0212373].
  %%CITATION = EPHJA,C28,515;%%

%\cite{Amsler:2008zz}
\bibitem{Amsler:2008zz}
  C.~Amsler {\it et al.}  [Particle Data Group],
  %``Review of particle physics,''
  Phys.\ Lett.\  B {\bf 667}, 1 (2008).
  %%CITATION = PHLTA,B667,1;%%



%\cite{Ball:2006eu}
\bibitem{Ball:2006eu}
  P.~Ball, G.~W.~Jones and R.~Zwicky,
  %``B --> V gamma beyond QCD factorisation,''
  Phys.\ Rev.\  D {\bf 75}, 054004 (2007)
  [arXiv:hep-ph/0612081].
  %%CITATION = PHRVA,D75,054004;%%



%\cite{Braun:2004vf}
\bibitem{Braun:2004vf}
  V.~M.~Braun and A.~Lenz,
  %``On the SU(3) symmetry-breaking corrections to meson distribution
  %amplitudes,''
  Phys.\ Rev.\  D {\bf 70}, 074020 (2004)
  [arXiv:hep-ph/0407282].
  %%CITATION = PHRVA,D70,074020;%%


%\cite{Ball:2005vx}
\bibitem{Ball:2005vx}
  P.~Ball and R.~Zwicky,
  %``SU(3) breaking of leading-twist K and K* distribution amplitudes: A
  %reprise,''
  Phys.\ Lett.\  B {\bf 633}, 289 (2006)
  [arXiv:hep-ph/0510338].
  %%CITATION = PHLTA,B633,289;%%


%\cite{Ball:2006nr}
\bibitem{Ball:2006nr}
  P.~Ball and R.~Zwicky,
  %``|V(td)/V(ts)| from B --> V gamma,''
  JHEP {\bf 0604}, 046 (2006)
  [arXiv:hep-ph/0603232].
  %%CITATION = JHEPA,0604,046;%%

%\cite{Ball:2007rt}
\bibitem{Ball:2007rt}
  P.~Ball and G.~W.~Jones,
  %``Twist-3 distribution amplitudes of K* and Phi mesons,''
  JHEP {\bf 0703}, 069 (2007)
  [arXiv:hep-ph/0702100].
  %%CITATION = JHEPA,0703,069;%%



%\cite{Ali:2007ff}
\bibitem{Ali:2007ff}
   A.~Ali, et al.,
  %``Charmless non-leptonic B/s decays to P P, P V and V V final states in the
  %pQCD approach,''
  Phys.\ Rev.\  D {\bf 76}, 074018 (2007)
  [arXiv:hep-ph/0703162].
  %%CITATION = PHRVA,D76,074018;%%

 C.D. Lu, Talk given at 4th International Workshop on the CKM
 Unitarity Triangle (CKM 2006), Nagoya, Japan, 12-16 Dec 2006 and
 talk given at 42nd Rencontres de Moriond on QCD and Hadronic
 Interactions, La Thuile, Italy, 17-24 Mar 2007, arXiv:0705.1782
 [hep-ph]

  %%CITATION = HEP-PH/0703162;%%





%\cite{Li:1994iu}
\bibitem{Li:1994iu}
  H.~n.~Li and H.~L.~Yu,
  %``Perturbative QCD Analysis Of B Meson Decays,''
  Phys.\ Rev.\  D {\bf 53}, 2480 (1996)
  [arXiv:hep-ph/9411308].
  %%CITATION = PHRVA,D53,2480;%%


%\cite{Li:2001ay}
\bibitem{Li:2001ay}
  H.~n.~Li,
  %``Threshold resummation for B meson decays,''
  Phys.\ Rev.\  D {\bf 66}, 094010 (2002)
  [arXiv:hep-ph/0102013].
  %%CITATION = PHRVA,D66,094010;%%


%\cite{Li:2002mi}
\bibitem{Li:2002mi}
  H.~n.~Li and K.~Ukai,
  %``Threshold resummation for nonleptonic B meson decays,''
  Phys.\ Lett.\  B {\bf 555}, 197 (2003)
  [arXiv:hep-ph/0211272].
  %%CITATION = PHLTA,B555,197;%%

%
%
%%\cite{DelDebbio:1997kr}
\bibitem{DelDebbio:1997kr}
  L.~Del Debbio, J.~M.~Flynn, L.~Lellouch and J.~Nieves  [UKQCD
                  Collaboration],
  %``Lattice-constrained parametrizations of form factors for semileptonic  and
  %rare radiative B decays,''
  Phys.\ Lett.\  B {\bf 416}, 392 (1998)
  [arXiv:hep-lat/9708008].

%\cite{Cheng:2003sm}
\bibitem{Cheng:2003sm}
  H.~Y.~Cheng, C.~K.~Chua and C.~W.~Hwang,
  %``Covariant light-front approach for s-wave and p-wave mesons: Its
  %application to decay constants and form factors,''
  Phys.\ Rev.\  D {\bf 69}, 074025 (2004)
  [arXiv:hep-ph/0310359].
  %%CITATION = PHRVA,D69,074025;%%




%\cite{Ball:2004rg}
\bibitem{Ball:2004rg}
  P.~Ball and R.~Zwicky,
  %``B/(d,s) --> rho, omega, K*, Phi decay form factors from light-cone sum
  %rules revisited,''
  Phys.\ Rev.\  D {\bf 71}, 014029 (2005)
  [arXiv:hep-ph/0412079].
  %%CITATION = PHRVA,D71,014029;%%
%


%\cite{Becirevic:2006nm}
\bibitem{Becirevic:2006nm}
  D.~Becirevic, V.~Lubicz and F.~Mescia,
  %``An estimate of the B --> K* gamma form factor,''
  Nucl.\ Phys.\  B {\bf 769}, 31 (2007)
  [arXiv:hep-ph/0611295].
  %%CITATION = NUPHA,B769,31;%%


%\cite{Lu:2007sg}
\bibitem{Lu:2007sg}
  C.~D.~Lu, W.~Wang and Z.~T.~Wei,
  %``Heavy-to-light form factors on the light cone,''
  Phys.\ Rev.\  D {\bf 76}, 014013 (2007)
  [arXiv:hep-ph/0701265].
  %%CITATION = PHRVA,D76,014013;%%


%\cite{Deandrea:1998ww}
\bibitem{Deandrea:1998ww}
  A.~Deandrea, R.~Gatto, G.~Nardulli and A.~D.~Polosa,
  %``Semileptonic B --> rho and B --> a1 transitions in a quark-meson  model,''
  Phys.\ Rev.\  D {\bf 59}, 074012 (1999)
  [arXiv:hep-ph/9811259].
  %%CITATION = PHRVA,D59,074012;%%


%\cite{Scora:1995ty}
\bibitem{Scora:1995ty}
  D.~Scora and N.~Isgur,
  %``Semileptonic meson decays in the quark model: An update,''
  Phys.\ Rev.\  D {\bf 52}, 2783 (1995)
  [arXiv:hep-ph/9503486].
  %%CITATION = PHRVA,D52,2783;%%

%\cite{Isgur:1988gb}
\bibitem{Isgur:1988gb}
  N.~Isgur, D.~Scora, B.~Grinstein and M.~B.~Wise,
  %``Semileptonic B and d Decays in the Quark Model,''
  Phys.\ Rev.\  D {\bf 39}, 799 (1989).
  %%CITATION = PHRVA,D39,799;%%

\bibitem{ckm}
CKMfitter Group (J. Charles et al.), Eur. Phys. J. C41, 1-131 (2005)
[hep-ph/0406184], updated results and plots available at:
http://ckmfitter.in2p3.fr



%\cite{Aliev:1999mx}
\bibitem{Aliev:1999mx}
  T.~M.~Aliev and M.~Savci,
  %``Semileptonic B --> a1 l nu decay in {QCD},''
  Phys.\ Lett.\  B {\bf 456}, 256 (1999)
  [arXiv:hep-ph/9901395].
  %%CITATION = PHLTA,B456,256;%%

%\cite{Lee:2006qj}
\bibitem{Lee:2006qj}
  J.~P.~Lee,
  %``Radiative B --> K(1) decays in the light-cone sum rules,''
  Phys.\ Rev.\  D {\bf 74}, 074001 (2006)
  [arXiv:hep-ph/0608087].
  %%CITATION = PHRVA,D74,074001;%%


\bibitem{Yang:BtoAinLCSR}
 K.C. Yang, Phys. Rev. D 78, 034018 (2008).

%\cite{Wang:2008bw}
\bibitem{Wang:2008bw}
  Z.~G.~Wang,
  %``Analysis of the $B\to a_1(1260)$ form-factors with light-cone QCD sum
  %rules,''
  Phys.\ Lett.\  B {\bf 666}, 477 (2008)
  [arXiv:0804.0907 [hep-ph]].
  %%CITATION = PHLTA,B666,477;%%


 %\cite{Cheng:2004yj}
\bibitem{Cheng:2004yj}
  H.~Y.~Cheng and C.~K.~Chua,
  %``Covariant light-front approach for B --> K* gamma, K(1) gamma, K(2)*  gamma
  %decays,''
  Phys.\ Rev.\  D {\bf 69}, 094007 (2004)
  [arXiv:hep-ph/0401141].
  %%CITATION = PHRVA,D69,094007;%%


%\cite{Aubert:2006dd,Aubert:2007xd}
\bibitem{Aubert:2006dd}
  B.~Aubert {\it et al.}  [BABAR Collaboration],
  %``Observation of B0 meson decay to a1(1260)+- pi-+,''
  Phys.\ Rev.\ Lett.\  {\bf 97}, 051802 (2006)
  [arXiv:hep-ex/0603050].
  %%CITATION = PRLTA,97,051802;%%

%\cite{:2007jn}
\bibitem{:2007jn}
  K.~Abe {\it et al.}  [Belle Collaboration],
  %``Measurement of the Branching Fraction for B^0 --> a1(1260)+-pi-+ with 535
  %Million BBbar Pairs,''
  arXiv:0706.3279 [hep-ex].
  %%CITATION = ARXIV:0706.3279;%%

%\cite{Aubert:2007xd}
\bibitem{Aubert:2007xd}
  B.~Aubert {\it et al.}  [The BABAR Collaboration],
  %``Observation of B-meson decays to b_1 pi and b_1 K,''
  Phys.\ Rev.\ Lett.\  {\bf 99}, 241803 (2007)
  [arXiv:0707.4561 [hep-ex]].
  %%CITATION = PRLTA,99,241803;%%


%\cite{Wang:2008hu}
\bibitem{Wang:2008hu}
  W.~Wang, R.~H.~Li and C.~D.~Lu,
  %``What can we learn from $B\to a_1(1260)(b_1(1235))\pi(K)$ decays?,''
  Phys.\ Rev.\  D {\bf 78}, 074009 (2008)
  [arXiv:0806.2510 [hep-ph]].
  %%CITATION = PHRVA,D78,074009;%%



%\cite{Bauer:2001cu}
\bibitem{Bauer:2001cu}
  C.~W.~Bauer, D.~Pirjol and I.~W.~Stewart,
  %``A proof of factorization for B --> D pi,''
  Phys.\ Rev.\ Lett.\  {\bf 87}, 201806 (2001)
  [arXiv:hep-ph/0107002].
  %%CITATION = PRLTA,87,201806;%%


%
%
%\bibitem{exp}M. Nakao, et. al., Belle Collaboration, Phys. Rev. D69,
%112001 (2004); B. Aubert, et al., BaBar collaboration, Phys. Rev.
%D70, 112006 (2004)
%%\cite{Beneke:1998sy}
%
%
%
%1614;%%
%
%%\cite{Li:2006jv}
%\bibitem{Li:2006jv}
%  H.~n.~Li and S.~Mishima,
%  %``Penguin-dominated B --> P V decays in NLO perturbative QCD,''
%  Phys.\ Rev.\  D {\bf 74}, 094020 (2006)
%  [arXiv:hep-ph/0608277].
%  %%CITATION = PHRVA,D74,094020;%%
%
%%\cite{Kwon:2004ri}
%\bibitem{Kwon:2004ri}
%  Y.~J.~Kwon and J.~P.~Lee,
%  %``Implications of the first observation of B --> K1 gamma,''
%  Phys.\ Rev.\  D {\bf 71}, 014009 (2005)
%  [arXiv:hep-ph/0409133].
%  %%CITATION = PHRVA,D71,014009;%%
%
%
%%\cite{Jamil Aslam:2005mc}
%\bibitem{Jamil Aslam:2005mc}
%  M.~Jamil Aslam and Riazuddin,
%  %``Branching ratio for B --> K(1) gamma decay in next-to-leading order in
%  %LEET,''
%  Phys.\ Rev.\  D {\bf 72}, 094019 (2005)
%  [arXiv:hep-ph/0509082].
%  %%CITATION = PHRVA,D72,094019;%%
%
%%\cite{Aslam:2006vh}
%\bibitem{Aslam:2006vh}
%  M.~J.~Aslam,
%  %``Annihilation contributions in B --> K(1) gamma decay in next-to-leading
%  %order in LEET and CP-asymmetry,''
%  Eur.\ Phys.\ J.\  C {\bf 49}, 651 (2007)
%  [arXiv:hep-ph/0604025].
%  %%CITATION = EPHJA,C49,651;%%
%
%%\cite{Jamil Aslam:2006bw}
%\bibitem{Jamil Aslam:2006bw}
%  M.~Jamil Aslam and Riazuddin,
%  %``Branching ratio and CP-asymmetry for B --> 1(1)P(1) gamma decays,''
%  Phys.\ Rev.\  D {\bf 75}, 034004 (2007)
%  [arXiv:hep-ph/0607114].
%  %%CITATION = PHRVA,D75,034004;%%
%

\bibitem{pqcd}
  H.~n.~Li,
  %``QCD aspects of exclusive B meson decays,''
  Prog.\ Part.\ Nucl.\ Phys.\  {\bf 51}, 85 (2003)
  [arXiv:hep-ph/0303116];
  %%CITATION = PPNPD,51,85;%%

  C.~D.~Lu,
  %``QCD in hadronic B decays,''
  Mod.\ Phys.\ Lett.\  A {\bf 22}, 615 (2007)
  [arXiv:0706.0589 [hep-ph]].
  %%CITATION = MPLAE,A22,615;%%



\end{thebibliography}
\end{document}